\documentclass{aa}  
\pdfoutput=1
\usepackage{graphicx}
\usepackage{amsmath}
\usepackage{txfonts}
\usepackage[]{natbib} 
\newcommand\lmbc{\lambda_c}

\begin{document}

\title{From molecules to Young Stellar Clusters: the star formation cycle across the M33 disk}

\titlerunning{From molecules to YSC in M33 }  
 
\author{Edvige Corbelli
       \inst{1}
	\and 
           Jonathan Braine
       \inst{2}
	\and
	   Rino Bandiera
         \inst{1} 
	\and 
	   Nathalie Brouillet
       \inst{2}
	\and
           Fran\c coise Combes
        \inst{3}
	\and 
	   Clement Druard
        \inst{2}
	\and 
           Pierre Gratier
        \inst{2}
	\and 
	   Jimmy Mata
        \inst{2}
	\and
           Karl Schuster
	\inst{4}
	\and 
           Manolis Xilouris
        \inst{5}
	 \and
	   Francesco Palla\thanks{Deceased}
         \inst{1}
		 }

   \institute{INAF-Osservatorio Astrofisico di Arcetri, Largo E. Fermi, 5,
             50125 Firenze, Italy\\
             \email{edvige@arcetri.astro.it,bandiera@arcetri.astro.it} 
             \and 
             Laboratoire d'Astrophysique de Bordeaux, Univ. Bordeaux, CNRS, B18N, 
                   all\'ee Geoffroy Saint-Hilaire, 33615 Pessac, France\\
                   \email{jonathan.braine@u-bordeaux.fr,
                   nathalie.brouillet@u-bordeaux.fr,clement.druard@gmail.com,
			pierre.gratier@u-bordeaux.fr,jimmy.mata@hotmail.fr}              	     		
             \and
             Observatoire de Paris, LERMA (CNRS: UMR 8112), 61 Av. de l'Observatoire, 75014, Paris, France\\ 
	     \email{francoise.combes@obspm.fr}
	     \and
	     Institut de Radioastronomie Millim\'etrique 
             300 rue de la Piscine, Domaine Universitaire 
             38406 Saint Martin d'H\`eres, France
	     \email{mailto:schuster@iram.fr}
	     \and
              Institute for Astronomy, Astrophysics, Space Applications $\&$ Remote Sensing, National 
              Observatory of Athens, P. Penteli, 15236, Athens, Greece\\
	      \email{xilouris@astro.noa.gr}
              }

   \date{Received .....; accepted ....}

 \abstract
   {}
   {We study the association between Giant Molecular Clouds (GMCs) and Young Stellar Cluster Candidates (YSCCs),
   to shed light on the time evolution of local star formation episodes in the nearby galaxy M33. }
   {The CO (J=2-1) IRAM-all-disk survey was used to identify and classify 566 GMCs with masses between 2$\times 10^4$ 
   and 2$\times 10^6$ M$_\odot$ across the whole star forming disk of M33. In the same area, there are 630 YSCCs, identified  using Spitzer-24~$\mu$m data.  
   Some YSCCs are embedded star-forming sites while the majority have GALEX-UV and H$\alpha$ counterparts with estimated cluster masses and ages. }
   {The GMC classes correspond to different cloud evolutionary stages: inactive clouds are 32$\%$ of the total, classified clouds with
   embedded and exposed star formation are 16$\%$ and 52$\%$ of the total respectively. Across the regular southern spiral arm, inactive clouds 
   are preferentially located 
   in the inner part of the arm,  possibly  suggesting a triggering of star formation as the cloud crosses the arm. 
   The spatial correlation between YSCCs and GMCs 
   is  extremely  strong, with a typical  separation  of 17~pc,  less than half the CO(2--1) beamsize,  illustrating the  remarkable physical 
   link between the two populations.   GMCs and YSCCs  follow the HI filaments, except in
   the outermost regions where the survey finds fewer GMCs than YSCCs  likely due to undetected, low CO-luminosity clouds. The distribution of 
   the non-embedded YSCC ages peaks around  5~Myrs with only a few being as old as  8--10~Myrs. These age estimates together with the  
   number of GMCs in the various evolutionary stages  lead us to conclude that  14~Myrs is a typical lifetime of a GMC in M33, prior to  
   cloud dispersal. The inactive and embedded phases are short, lasting about 4 and 2~Myrs respectively.
   This underlines that embedded YSCCs rapidly break out from the clouds and become partially visible in H$\alpha$ 
   or UV long before cloud dispersal.  
   }
  {}

   \keywords{Galaxies: individual (M\,33) --
             Galaxies: ISM --
                      }
   \maketitle
 
\section{Introduction}

The formation of giant molecular clouds (hereafter GMCs) in the 
bright disks of spiral galaxies requires the onset of instabilities and  
the ability of the gas to cool and fragment.  Within  gravitationally unstable clouds, the process of fragmentation continues to  smaller scales, yielding a distribution of  
clumps  and prestellar cores which later collapse to form stars.  The evolution of  these clouds is then driven by the
Young Stellar Clusters (hereafter YSC) which have formed.  As the YSC evolves, it could disrupt the cloud or trigger  new episodes of star formation.  Our Galaxy is the
natural laboratory where these processes have been studied in detail because observations can be carried out with an unbeatable  spatial resolution.
Galactic observations suffer from limitations  due to the fact that we reside within the
star forming disk. Moreover, galaxies with different masses, morphologies, metal contents and in different environment or cosmic time, might
transform the gas into stars over different timescales and with different efficiencies. The molecular gas fragments, from GMCs to protostellar clumps, 
do not necessarily follow the same mass spectrum as Galactic clouds nor share their characteristics.
These considerations have triggered great interest in studying 
 molecular clouds in nearby galaxies, with the support of  millimeter telescopes which are steadily improving in resolution and sensitivity (e.g. {\it ALMA,NOEMA}).
 Validating a cloud formation and evolution model requires an unbiased survey of molecular clouds in a galaxy and of star-forming sites.
 
 All-disk-surveys of the $^{12}$CO J=1-0 or J=2-1 line  emission  in nearby galaxies, are the most commonly used to provide  a census of molecular complexes  
 down to a certain sensitivity limit. The Large Magellanic Cloud in the southern hemisphere, and M33 in the northern hemisphere,  have been targets
 of  several observing campaigns of molecular gas emission,  as they are nearby,  gas rich and with active star formation \citep{1999PASJ...51..745F,2008ApJS..178...56F,
 2003ApJS..149..343E,2004ApJ...602..723H,2010A&A...522A...3G,2014A&A...567A.118D}. 
 The NANTEN group studied the LMC and identified 168 clouds \citep{2001PASJ...53..971M, 2001PASJ...53L..41F,2008ApJS..178...56F}. \citet{1990ApJ...363..435W}
 surveyed the inner 2~kpc of M33 and detected 38 GMC with a 7" synthesized beam.
 The $^{12}$CO J=1-0 survey of the M33 disk with the BIMA interferometer with a 13" synthesized beam,  has provided a catalogue of 148 clouds out to R=6~kpc,
 complete down to 1.5$\times$10$^5$~M$_\odot$ \citep{2003ApJS..149..343E,2003ApJ...599..258R,2007ApJ...661..830R}.  
 \citet{2012A&A...542A.108G} analyzed the IRAM CO J=2-1 survey of a large area of the   
 disk of M33 and catalogued 337 GMCs.  The star forming disk 
 of M33 was observed in the CO J=1-0 and J=3-2 line  with an angular 
 resolution of 25"  by \citet{2012ApJ...761...37M} who identified 71 GMCs, 70 of which were already in the \citet{2007ApJ...661..830R} or \citet{2012A&A...542A.108G} catalogues.
The data used here are from the deep CO(J=2-1) whole-disk survey carried out with the IRAM-30m telescope \citep{2014A&A...567A.118D}  at 12$"$ resolution (49pc) from which 566 clouds 
were identified. 
    
Molecular clouds are not located around or very close to  optically visible stellar  clusters, as it will be shown in this paper.
In the early phases of star formation (hereafter SF), protostars and stars are embedded in the gas and can be detected via imaging with infrared telescopes due to the high extinction 
provided by the molecular material.  The {\it Spitzer Space Telescope} has surveyed the LMC and M33 in the Mid-Infrared (hereafter MIR)  with sufficient spatial resolution 
as to provide a 
detailed view of where  the hot dust emission is located.  Emission from hot dust, typically detected at $24 \mu$m, is an excellent tracer of star forming sites where  massive or 
intermediate mass stars have formed. 
The detection of protostars and of the earliest phases of SF is not yet feasible in M33 since it requires  radio and far-infrared surveys with  far 
higher resolution than what is currently available.
Searches for embedded star forming sites in M33 are limited
to less compact sites,  i.e. to stellar clusters within clouds which are visible at 24$\mu$m in the {\it Spitzer} survey of the 
whole galaxy \citep{2007A&A...476.1161V,2009A&A...493..453V}.  A catalogue of  MIR  emitting sites in M33 is  now available
\citep{2011A&A...534A..96S} and  a list of candidate star-forming sites has been  selected  to investigate their distribution in mass and  spatially  in the disk of M33.
After formation, stars may clear some of the gas such that, depending on viewing angle,  optical counterparts to the MIR emission can be found.  The YSC is still compact and may still 
suffer from modest extinction such that individual members cannot be resolved.  
In this context, we study the relationship between Young Stellar Cluster Candidates  (hereafter YSCCs, referred here only to sources which have been selected via their MIR  emission
and are listed in Table 6),  and the molecular cloud population  
in  M33. A detailed analysis of the spatial correlation between GMCs and YSCCs is carried out across the whole star-forming disk of M33. 
Using essentially the same classification scheme as in \citet{2012A&A...542A.108G}, we define classes of 
GMCs in terms of their star formation. 
Combined with the classification and age determination of YSCCs, it is possible to estimate durations for the various phases of the star formation 
process and the GMC lifetime in this nearby galaxy. This is important in order to link the physical conditions  within GMCs to the processes 
which regulate star formation, its efficiency, and possible time variations \citep{2007ApJ...668.1064E,2011ApJ...729..133M}.    

The paper plan is the following. In Section 2 we describe the new GMC catalogue and introduce the cloud classification scheme. In Section 3  we describe the MIR-source
catalogue and introduce the YSCC classification scheme.  In Section 3 we also discuss the association between the GMCs and  YSCCs 
and in Section 4 the association between GMCs and  optically visible YSCs and other sources related to the SF cycle. The catalogs of GMCs and YSCCs and 
their classification are provided in the on-line Tables. 
The properties  of molecular cloud classes and of YSCC classes are presented in Sections 5 and 6. Molecular cloud lifetimes  across 
the M33 disk are  analyzed  in Section 7. Section 8 summarizes the main results.
 
\section{The molecular cloud population}

\begin{figure} 
\includegraphics[width=9cm]{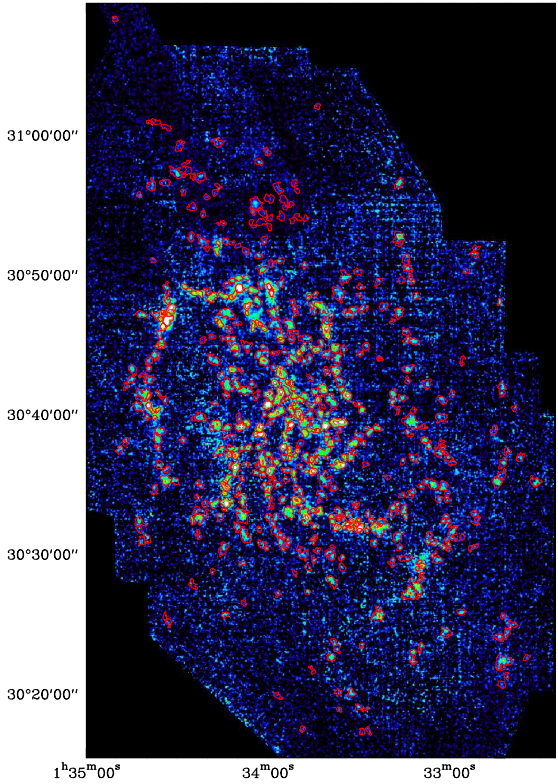}
\caption{Molecular clouds in the catalogue  are  plotted over the CO J=2-1 integrated intensity emission map. The map rms  is 0.2~K~km~s$^{-1}$ 
and the intensity scale reaches a maximum of 4~K~km~s$^{-1}$ in antenna temperature units.
The thick red contour of each cloud corresponds to the deconvolved size 
and shape as determined by CPROPS  \citep{2006PASP..118..590R}. }
\label{COclo} 
\end{figure}

M33 is a low-luminosity spiral galaxy, the third most luminous member of the Local Group, with a well determined distance 
D=840~kpc \citep{1991ApJ...372..455F,2013ApJ...773...69G}. The GMC catalogue is a product of the  IRAM-30m all-disk  CO J=2-1 survey  of M33 presented in
 \citet{2014A&A...567A.118D} using a modified version of the CPROPS package, originally developed by \citet{2006PASP..118..590R}, which is described in detail by 
\citet{2012A&A...542A.108G}. The CO datacube has a spatial resolution of 12~arcsec which, for the adopted distance of M33, corresponds to a physical scale of 49~pc
(hence, GMCs are not well resolved). The spectral resolution of the datacube is 2.6~km~s$^{-1}$, and the pixel size is 3~arcsec (i.e. 12~pc). 
CPROPS identifies continuous regions of CO emission in the datacube and details on our use can be found in \citet{2012A&A...542A.108G}.
In Figure \ref{COclo} we show the location of the clouds on the CO (J=2-1) map. 

The cloud mass is computed either converting the total CO line luminosity of the cloud into mass, referred to as 
luminous mass, or using the virial relation, referred to as  virial mass.
The luminous mass of the GMC, M$_{H_2}$, is computed for a hydrogen fraction $f_h$ of 73$\%$, using a CO-to-H$_2$ conversion factor
X=$N(H_2)/I_{CO(1-0)}=4\times 10^{20}$~cm$^{-2}$/(K~km~s$^{-1}$) \citep{2016arXiv160903791G}, twice the standard Galactic value, and an 
intrinsic line ratio $R^{2-1}_{1-0}=I_{2-1}/I_{1-0}= 0.8$ \citep{2014A&A...567A.118D}. The luminous mass is a function
of  the  CO (J=2-1) line luminosity, L$_{CO}$, which is the CO J=2-1 line intensity integrated over the cloud. 
Another way of estimating cloud masses is to assume virial equilibrium.
The virial mass, M$_{H_2}^{vir}$,  
is a function of the deconvolved effective cloud radius, r$_e$, and of the CO (J=2-1) line dispersion \citep{1987ApJ...319..730S,2006PASP..118..590R}. 
The line dispersion  is measured by fitting a gaussian function to the cloud integrated line profile. If $\Delta V_{FWHM}^{gau}$ is the full
width half maximum of the fitted line profile, corrected for the finite channel width (by subtracting in quadrature 2.6~km~s$^{-1}$),
then $\sigma_v^{gau} = \Delta V_{FWHM}^{gau}/\sqrt{8ln2}$.
We apply the following relations which include chemical elements heavier than hydrogen:

\begin{eqnarray}
{{\hbox{M}}_{H_2}\over M_\odot} & =  & {19.1\over  R^{2-1}_{1-0}}\ \  {{\hbox {X}} \over 4\times 10^{20}} {{2 \ m_p} \over f_h} 
\ {{\hbox{L}}_{CO(2-1)}\over K~km~s^{-1}~pc^2}    \\
& = & 10.9 \ {{\hbox{L}}_{CO(2-1)}\over K~km~s^{-1}~pc^2} 
\end{eqnarray}

\begin{equation}
 {{\hbox {M}}_{H_2}^{vir}\over M_\odot}= 1040\  
{{\hbox{r}}_e \over pc} \Bigl({\sigma_v^{gau} \over km~s^{-1}}\Bigr)^2
\end{equation}

Typical cloud linewidths are of order 6-10~km~sec$^{-1}$ 
with $\sigma_v\sim$3-4~km~sec$^{-1}$. Cloud radii vary between 10 and 100~pc, typical of GMCs 
and complexes. However, given the spatial resolution of the survey, cloud radii have large uncertainties
and in some cases they might be overestimated. As a consequence virial masses often turn out to
be larger than luminous masses, and scaling relations may apply only to smaller clouds within a GMC complex. 
We shall use the luminous mass definition when we refer to the GMC mass, unless stated otherwise.
The algorithm finds 566 GMCs (see Fig. \ref{COclo}) with luminous masses between 2$\times 10^4$ and 2$\times 10^6$ M$_\odot$ and virial masses between 2$\times 10^4$ and 
6$\times 10^6$ M$_\odot$. 

\begin{figure} 
\includegraphics[width=9cm]{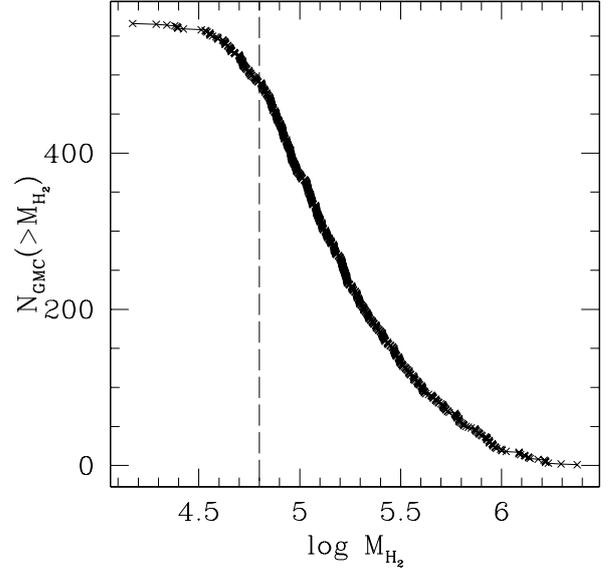}
\caption{The cumulative distribution of the GMC luminous masses. N$_{GMC}$ is the number of GMCs with mass greater than M$_{H_2}$.
The dashed line has been placed at the estimated survey completeness limit.}
\label{cum} 
\end{figure}

The completeness limit for the luminous masses is about 2/3 of what has been estimated  by \citet{2012A&A...542A.108G} due to the lower rms noise of the  
\citet{2014A&A...567A.118D}  full-disk survey and to the revision of telescope
efficiency. Hence we estimate a completeness limit of 5700 K~km~s$^{-1}$~pc$^2$. Given the assumed CO-to-H$_2$ conversion factor  and J=2-1/J=1-0 line ratio,
this corresponds to a total cloud mass completeness limit of $6.3 \times 10^4$~M$_\odot$, including He.
Figure \ref{cum} shows the cumulative distribution of the 566 GMC masses.
 
\begin{figure} 
\includegraphics[width=9cm]{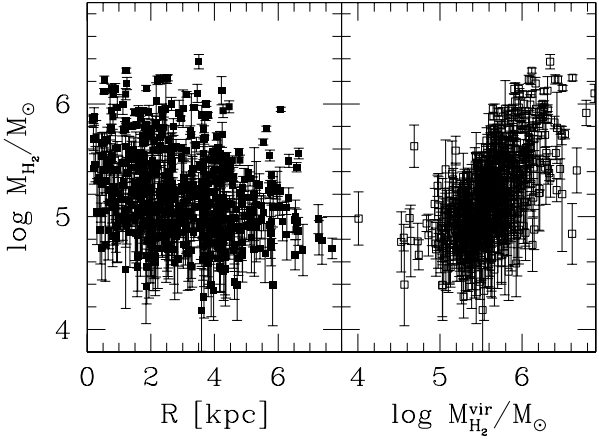}
\caption{In the left panel the cloud luminous mass M$_{H_2}$ is plotted as a function of the galactocentric radius in kpc.
In the right panel we show the luminous versus the virial masses of the clouds.  }
\label{mlum} 
\end{figure}

In the left panel of Figure \ref{mlum} we show the radial distribution of the luminous masses of the GMCs
while in the right panel we plot the luminous and virial masses of the clouds. The
average luminous mass decreases with radius because the CO cloud luminosity is  a decreasing function of galactocentric radius 
\citep{2012A&A...542A.108G}. There are fewer massive cloud complexes beyond 4.5~kpc.
The virial mass shows a marginal dependence on galactocentric radius because   the velocity dispersion  (line width) is weakly anti-correlated with
the galactocentric distance of the cloud \citep{2012A&A...542A.108G,2010A&A...520A.107B}. Clearly the luminous and virial masses are correlated
even though the dispersion is non-negligible.  The right panel of Figure~\ref{mlum} seems to suggest that  GMCs may simply be 
gravitationally bound entities, not necessarily in virial equilibrium.

\subsection{Cloud Classification}

\begin{figure} 
 \includegraphics[width=7cm]{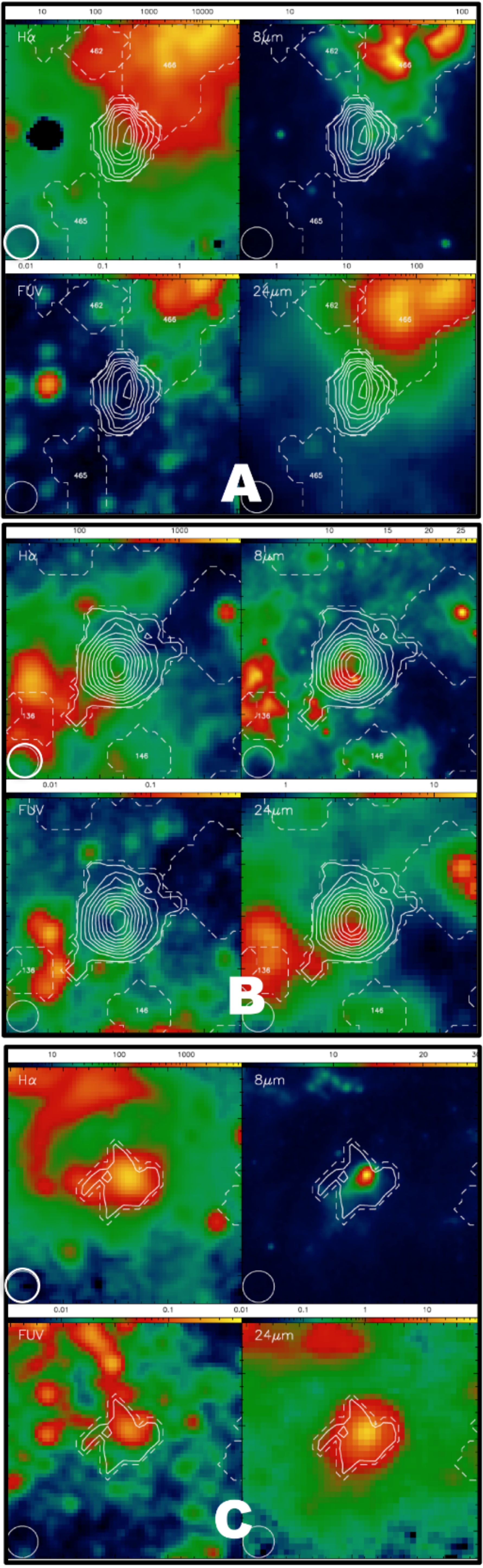}
\caption{ Four images of 3 selected areas of the M33 disk used to classify 3 GMCs: cloud 461 as inactive or A-type (top), cloud 147 with embedded SF or B-type (middle), 
and cloud 15 with exposed SF or C-type (bottom).
The CO J=2-1 integrated intensity contours for each GMC (solid white line, 
first contour at 80~mK~km~s$^{-1}$) are plotted on maps of H$\alpha$ (upper left panel), Spitzer 8~$\mu$m (upper right panel), 
GALEX FUV (lower left panel), and Spitzer 24~$\mu$m (lower right panel).  
The CO J=2-1 beam is plotted in the lower left corner of each panel and it is marked with a thicker line in the upper left panel.
 }
\label{multiCO} 
\end{figure}

Molecular clouds are classified in three broad categories -- clouds without obvious star formation (A), clouds with embedded star formation (B) and clouds with 
exposed star formation (C). Clouds with embedded or exposed SF are identified from the presence of emission at 8 or 24 $\mu$m, that in C-type clouds is associated
with H$\alpha$ and often to Far-UV emission peaks, while B-type clouds have no optical counterpart. There are a few ambiguous cases which were classified D-type.  
This classification follows the spirit of the \citet{2012A&A...542A.108G} procedure in which maps at four wavelengths were made of each cloud and the region surrounding 
it (see catalog in \citet{2012A&A...542A.108G}) as displayed in Figure \ref{multiCO}.  The wavebands were chosen to probe a variety of optical depths, at a higher angular 
resolution than the CO data in order to locate the SF region within the molecular cloud.  

The 566 clouds were visually inspected using the maps as shown in Figure \ref{multiCO}.  Strict automatic criteria are extremely difficult to use in a reliable 
way as the general flux levels decrease greatly with galactocentric radius, making a common threshold impractical as one misses regions of star formation in 
the outer disk. Furthermore, crowded regions (center, arms) are difficult to analyse without visual inspection.  
It is not possible to know whether continuum emission is from the cloud or simply along the line of sight through the disk.  
When the emission is from a region near the center of the cloud, or there is line emission at the same location, the association is assumed to be real.
In Figure \ref{multiCO}, we show  examples of  the images used to classify each of the 566 GMCs. The CO J=2-1 integrated intensity contours for each GMC (solid white line, 
first contour at 80~mK~km~s$^{-1}$ and following stepped by 330~mK~km~s$^{-1}$) are plotted on maps of H$\alpha$ (upper left panel, units give emission measure in pc cm$^{-6}$), 
Spitzer 8~$\mu$m (upper right panel, MJy/sr), GALEX FUV (lower left panel, counts), and Spitzer 24~$\mu$m (lower right panel, MJy/sr). The  flux levels are different
for each image. The scaling used is plotted as a color bar at the top of each panel.  
One can see that for GMC 461 (upper four panels)
there are no visible sources in any band, and hence it has been classified as A-type. The GMC 147 (middle four panels) has a weak infrared source visible 
at 8 and 24$\mu$m with no H$\alpha$ or FUV emission peak at the MIR peak location and hence it has been classified as B-type. 
This cloud is located close to the M33 center, in a crowded region, and the weak MIR-source is not present in the \citet{2011A&A...534A..96S} catalogue discussed 
in the next Section. This underlines the need for visual inspection for a reliable cloud classification. At the  location of GMC 15 (bottom four panels) 
there is a source visible at all selected wavelengths, which corresponds to  source 8 in the catalogue of \citet{2011A&A...534A..96S} (see next Section)
and hence GMC 15 has been classified as C-type cloud.

A star forming region on the distant side of a cloud will be classified as B although an observer, in say M~31, might see the cloud as exposed (C). Thus, a 
cloud could be more evolved than its classification shows either due to geometry (the case above) or if only low-mass star formation is taking place.  
We estimate that the maps are sensitive enough to detect a single B0 main sequence star.  
There are some differences with respect to the  \citet{2012A&A...542A.108G} classification.  A slightly higher fraction of A clouds is found due to 
the classification of some clouds with very weak $and$ diffuse IR emission as A rather than B or C.  A slightly higher fraction of C-type clouds is also 
found because, with the idea that an apparently embedded cloud could be exposed from a different vantage point, we pushed ambiguous B/C cases into the C class.  
However, generally the agreement with  \citet{2012A&A...542A.108G} is excellent.

Table \ref{GMCclasses} summarizes the classification scheme and the number of GMCs in each class.
In Table~\ref{tabco} we list for each cloud the cloud type and the following properties: celestial coordinates, galactocentric radius R,  cloud deconvolved effective radius
r$_e$, CO(2-1) line velocity dispersion from CPROPS $\sigma_v$ and its uncertainty, line velocity dispersion from gaussian fit $\sigma_v^{gau}$ (corrected for finite channel width), 
CO luminous mass M$_{H_2}$ and its uncertainty, and virial mass from gaussian fit M$_{H_2}^{vir}$.  
The uncertianties on the velocity dispersion from CPROPS can be considered upper limits to the
uncertainties on the velocity dispersion from gaussian fit. Using these and the uncertainties on the cloud radius one can verify the large uncertainties on the virial mass estimates. 
There are no
estimates of $\sigma_v^{gau}$ and M$_{H_2}^{vir}$ when the gaussian fit to the cloud integrated profile results in a full width half maximum comparable to spectral resolution.
In addition we give  in Table \ref{GMCclasses} the number of the YSCC associated with the MIR-source and with the GMC, as described in the next 
Section, that lies within the 80~mK~km~s$^{-1}$ GMC boundary. Clouds of A-type should not host any source and in fact only 5 clouds in this category have a 
YSCC associated with them which lies at the cloud boundary. 
The cloud classification was done by seven testers without knowledge of the MIR-source position. At a later time the possible presence of sources at the GMC positions
has been checked by inspecting the whole M33 maps at 8~$\mu$m, 24~$\mu$m, H$\alpha$, and GALEX-FUV, each with a uniform contrast, and by analyzing statistically
the spatial correlation of catalogued YSCCs and GMCs (see next Section).

\begin{table}
\caption{GMC classification scheme} 
\centering                                       
\begin{tabular}{c c c }           
\hline\hline 
Class &   main properties &  $\#$GMCs  \\                  
\hline\hline 
A  &  inactive &  172   \\
B  &  embedded SF, MIR-sources only  &  87     \\
C  &  exposed SF, MIR+FUV+H$\alpha$ sources &  286    \\
D  &  ambiguous &   21   \\
\hline\hline 
\end{tabular}
\label{GMCclasses}
\end{table}

\section{Young Stellar Cluster Candidates in the disk of M33 and their association with GMCs}

In the star formation cycle, YSCs form out of molecular gas; as the cluster
evolves, stellar activity removes the molecular material and light starts to  escape  from the clouds. Eventually, shocks due to massive stars may compress
the nearby gas 
and trigger star formation anew. Before the cold gas is removed from the stellar birth place, newborn stars 
heat the dust in the surroundings with a consequent emission in the MIR. 
Therefore we expect the presence of YSCCs at the location of MIR-sources associated with 
the M33 disk and a spatial correlation between YSCCs and  star forming GMCs. 
The establishment of the association between YSCCs and GMCs is done following 3 
different methods with different levels of accuracy; each method is described in the following  subsections.
 
Using the Spitzer satellite 24$\mu$m data, \citet{2011A&A...534A..96S} have selected 915 MIR-sources in the area covered by the M33 disk. 
 Complementing the mid and far-infrared Spitzer data with UV data from the GALEX satellite  and with H$\alpha$ data, 
 it has been possible to build up the Spectral Energy Distribution (SED) for most of the sources.
  The optical images cover a smaller region than the Spitzer images and hence  for about 60  MIR-sources  at large 
 galactic radii the SED could not be derived.
  The presence in the sample of a few AGBs, the  weakness of the emission in the H$\alpha$ or UV bands  of some sources, and  some large photometric errors,
 has further limited to 648 the number of  MIR-sources  with available SEDs  \citep{2011A&A...534A..96S}. 
  We find that 738 out of the 915 MIR-sources catalogued by \citet{2011A&A...534A..96S} are within the area of the CO survey. 
 By visually inspecting these 738 sources in several bands and
 with available stellar catalogues we eliminated obvious AGB and Milky Way stars as well as background galaxies. The SDSS has been used to inspect the source
 optical morphology or  the photometric redshift  when the associated H$\alpha$ emission is weak,
 below $\sim 10^{36}$~erg~sec$^{-1}$. We have excluded sources with  a reliable redshift determination (with a $\chi^2<3$ as given by SDSS),  
 and a few MIR-sources with  X-ray counterparts  since these are probably background sources (QSOs, galaxies etc..).

We have a final sample of 630 MIR-sources which are strong candidates for being star forming sites over the area covered by the IRAM CO-all-disk survey and
we shall refer to the  young stellar clusters associated with these MIR-sources as YSCCs.
  The purpose of the identification of YSCCs in this paper is to associate them  with GMCs to study  cloud  and star 
 formation properties across M33. The YSCCs may have an optical counterpart, 
 or may be fully embedded, detected only in the infrared while stars are still forming. Soon after stars of moderate mass are born, the dust in the surrounding 
 molecular  material absorbs almost all the UV and optical emission of the recently born stars and re-emits the radiation in the Mid and Far-IR. Hence, MIR-sources without 
 optical or UV counterparts might indicate the presence of recently born stars still in their embedded phase. Furthermore  one has to bear in mind that small star 
 forming sites might be below the critical mass to fully populate the IMF and only occasionally form a massive stellar outlier with H$\alpha$ or FUV  luminosity 
 above the detection threshold. 
 This implies that for the purpose of this paper, i.e. association of YSCCs with molecular clouds, both MIR-sources  with and without UV or H$\alpha$
 counterparts are of interest. Ages and masses are available for 506 YSCCs with UV and H$\alpha$ emission, and have been determined by \citet{2011A&A...534A..96S}.

\subsection{The association between GMCs and YSCCs: filamentary structures across M33}

\begin{figure} 
\centerline{
\includegraphics[width=9cm]{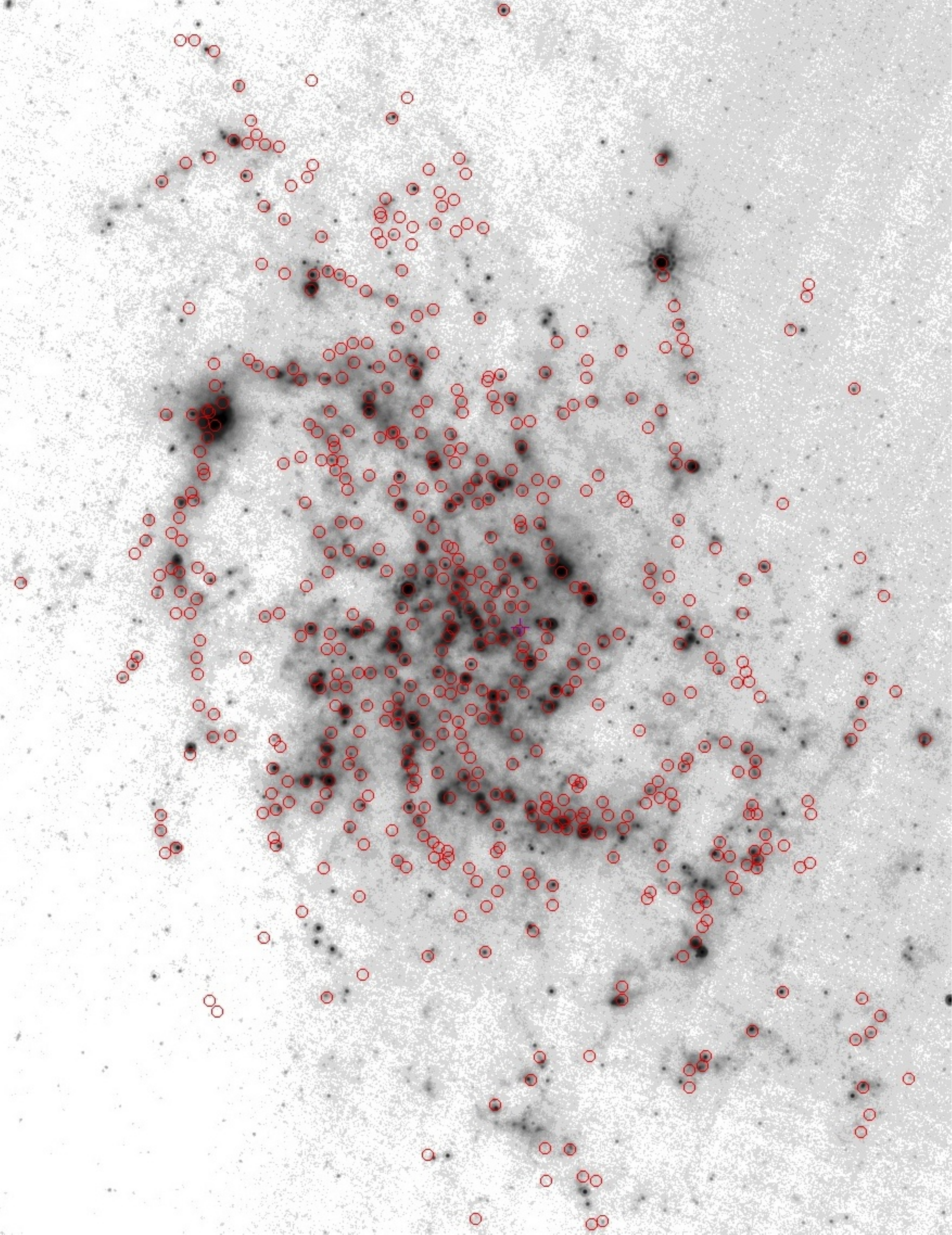}}
 \caption{Positions of CO-clouds (red circles) over the 24~$\mu$m Spitzer map. 
 }
\label{spitzer} 
\end{figure} 

In Figure \ref{spitzer} we plot the position of the 566 GMCs over the 24~$\mu$m image to show their large scale distribution over a MIR-Spitzer 
image of M33. Even though the correspondence between MIR-peaks, which are YSCCs, and GMCs is not a one-to-one correspondence, the majority of GMCs lie along 
filaments traced by the MIR emission. 
There are no GMCs in areas devoid of MIR emission. There are however some regions where MIR filaments are present but no GMCs have been found.
Similarly, some GMCs are present along tenuous and diffuse MIR filaments but don't overlap with emission peaks i.e. with compact sources as those detected by 
the \citet{2011A&A...534A..96S} extraction algorithm. Even using the Spitzer 8~$\mu$m map at 3~arcsec 
resolution  (better than the 6~arcsec resolution of 24~$\mu$m map)  some of these clouds seem associated only  with diffuse MIR emission.
 
\begin{figure} 
 \includegraphics[width=9cm]{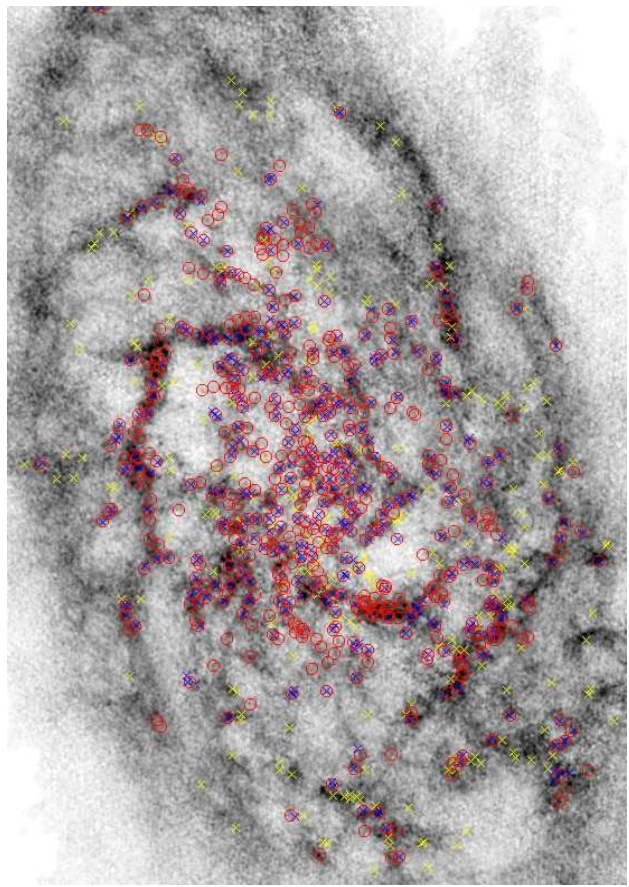}
 \caption{Positions of GMCs and YSCCs analyzed in this paper are plotted on top of the 21-cm map at 10~arcsec resolution.  
 Red  circles   indicate  the positions of GMCs, yellow and blue crosses indicate  the position of YSCCs not associated and associated with clouds
respectively.}
\label{HI} 
\end{figure}

 In Figure \ref{HI} we plot the GMC positions over the HI map at a spatial resolution of 10~arcsec, very similar to the CO map resolution.
 The 21-cm map, presented in \citet{2014A&A...572A..23C}  is obtained by combining  VLA and GBT data.  There is an extraordinary spatial 
 correspondence between the GMCs and the distribution of atomic hydrogen overdensities, underlined also by \citet{2003ApJS..149..343E}  and quantified by 
\citet{2012A&A...542A.108G}. This correspondence seems to weaken at large galactocentric radii. 
 Here, in fact, we  notice the presence of bright HI filaments  in areas devoid of GMCs.
 This may be due to a decrease of the CO J=2-1 line brightness far from the galaxy center
 because of CO dissociation, or 
  due to a gradient in metallicity \citep{2010A&A...512A..63M} or gas density  (which implies a lower CO J=2-1/J=1-0 line ratio).  
 Another possibility is that  fewer GMCs are formed in the absence of spiral arms. 
 Spiral arms  may favor the growth of GMCs by collisional aggregation of 
 smaller clouds. In the absence of the arms, only individual molecular clouds of lower mass and size than GMCs may be found, undetected by the survey 
 because of beam dilution.
 In the outer regions most of the CO J=2-1 emission is in fact  diffuse,  at the 12$"$ resolution of our CO data, which may be due to low mass clouds.
 For $R<4$~kpc most of the detected CO emission is due to GMCs in the catalogue \citep{Druardthesis}.   
 Furthermore, as pointed out already by \citet{2003ApJS..149..343E}, the presence of a
 high HI  surface density is a necessary condition but not a sufficient one for the formation of molecular clouds: the atomic gas might just not be converted 
 into molecules if hydrostatic pressure and  the dust content decrease as it happens going radially outwards in a spiral disk.   
 In Section~7 we will discuss further what can cause the drop in the number density of GMCs in the outer disk of M33  by examining the association of GMCs with YSCCs 
 and the GMC lifetime.

\subsection{The association between GMCs and YSCCs: a close inspection of cloud boundaries and the YSCC classification}

The spatial correspondence between the position of GMCs and that of YSCCs can be studied by an accurate inspection of the area covered by each
GMC. We start by searching for YSCCs which are within  1.5 cloud
radii of all GMCs listed in Table~\ref{tabco}. Since GMCs are often not spherical, we used a search radius larger than the cloud radius and subsequently we checked the 
association by inspecting visually the GMC contours drawn over the M33 {\it Spitzer} images at 8 and 24~$\mu$m (as for cloud classification). 
If in projection a YSCC and a GMC overlap we claim they are associated. 

We searched for optical or UV counterparts to YSCCs  by analyzing the H$\alpha$ and GALEX-FUV images of M33 and by checking the SDSS image at the location 
of each MIR-source. Taking into account whether a YSCC is associated with a  GMC  or not and whether it has or not an optical counterpart,
we place the 630 YSCCs into 4 different categories.
We describe these categories briefly below and discuss them in more detail in the following sections. 

\begin {itemize}
{\item class b}:   YSCCs, associated with GMCs, with no optical counterpart (unidentifiable in SDSS and with no H$\alpha$ emission)    
{\item class c}:   YSCCs with optical (SDSS and/or H$\alpha$) counterpart  
\begin {itemize}
{\item c1}:  YSCCs associated with GMCs with coincident H$\alpha$ and MIR emission peaks but FUV emission peaks are spatially shifted or absent 
{\item c2}:  YSCCs associated with GMCs with coincident H$\alpha$, FUV and MIR emission peaks  
{\item c3}:  YSCCs not associated with GMCs but star forming with optical and FUV counterparts; these often have weak H$\alpha$ emission 
\end {itemize}
{\item class d}:  YSCCs associated with GMCs which are ambiguous for b or c1/c2 class
{\item class e}:  YSCCs not associated with GMCs and with no H$\alpha$ emission and no or weak  red optical counterpart in SDSS; some FUV may be present 
\end {itemize}

The lowercase letters used for YSCC categories are such that if there are GMCs associated with them, these are mostly placed in the corresponding capital 
letter class. This is why we do not have a-type MIR-sources, because A-type GMCs are not associated with YSCCs. Similarly we do not have E-type GMCs because
the YSCCs in the class e  are not associated with GMCs.
After an automated search we inspected the images and checked that a YSCC associated with a GMC
lies within the cloud boundary and whether it has H$\alpha$, FUV,   and/or optical counterparts. We find that 243 YSCCs lie beyond any catalogued cloud borders and 104
of these have no H$\alpha$ counterpart and weak or no emission in the UV. We place these sources in the class e. Some of them  might not 
be star forming sites but foreground or background objects; however, in the class e we may also find some small, embedded star forming region whose associated 
GMC has a brightness below the survey detection threshold.  
We place the remaining 139 YSCCs not associated with GMCs in the c3 class. Optically these look like SF regions. Some might be YSC associated with smaller 
clouds which are not in the catalogue and others might be associated with more evolved sources whose original cold gas reservoir has been mostly dissipated. 
By looking at the ages and masses for c3-type YSCCs we find  ages  similar to those for c1- or c2-type YSCCs, while masses are smaller on average.
Hence it is likely that the majority of the YSCCs of class c3 are associated with molecular clouds of smaller mass, undetected by the survey.

We have 387 YSCCs (61$\%$ of the total) which have a high probability  of being linked to catalogued GMCs since they are within the cloud boundary. 
A few sources are associated   
with more than one cloud (since they are spatially overlapping with two clouds which are at different velocities or are at the boundary of two clouds). 
We classified 368 of these 387 YSCCs as  c1, c2 , or b-type according to the presence of an optical counterpart. We place the remaining 19 YSCCs, ambiguous 
between b- and c-type sources, in the  class d. 
We find an optical counterpart to the majority of YSCCs  associated with GMCs: 271 out of 368. Only 97 YSCCs, i.e.  26$\%$ of YSCCs associated with GMCs, 
do not have an optical counterpart and are candidates for being  YSCC still in their fully embedded phase. The 8 and 24~$\mu$m images of a b-type YSCC 
can be found in the central panels of Figure~\ref{multiCO}.  Each  of the 271 YSCC associated with a GMC has been 
classified as c1 or c2-type according to whether the FUV emission peak is absent/shifted with respect to the PSF of the 24~$\mu$m source or not.  
An example of a c2-type YSCC is shown at the center of the four bottom panels of Figure~\ref{multiCO} at various wavelengths. 
The summary of the YSCC classification is given in Table 2.

\begin{table}
\caption{YSCC classification scheme} 
\centering                                       
\begin{tabular}{c c c}           
\hline\hline 
Class & main properties & $\#$ YSCCs  \\                  
\hline\hline 
b    &  GMC, embedded SF   &  97   \\
c    &  exposed SF (c1+c2+c3)    & (410)      \\
c1   &  GMC, H$\alpha$, MIR coincident   &  55 \\
c2   &  GMC, H$\alpha$, MIR, FUV coincident & 216  \\
c3   &  no GMC, with FUV, weak H$\alpha$    & 139  \\ 
d    &  GMC, ambiguous                   & 19  \\
e    &  no GMC, no optical counterpart &  104 \\  
\hline\hline 
\end{tabular}
\label{MIRclasses }
\end{table}

The 630 YSCCs which are within the CO-all-disk survey map boundary are listed in Table~\ref{tabmir} with the type (from b to e) as described in the previous paragraph,
and the number of the GMCs associated with them, if any. In this case, we also list the corresponding cloud class (A, B, C, D).  
In addition, for each YSCC we give the celestial coordinates, the bolometric, total infrared, FUV and H$\alpha$ luminosities, the estimated mass and age, 
the visual extinction, the galactocentric radius, the source size, and its flux at 24~$\mu$m. The estimates of all  quantities given in Table~\ref{tabmir} and
their uncertainties are discussed in \citet{2011A&A...534A..96S}. Photometric errors on source luminosities are smaller than 0.1~dex. However, since for a source of a 
given size we perform surface
photometry with a fixed aperture at all wavelengths, the errors due to nearby source contamination  can increase the uncertainties in crowded field 
(FUV in center and spiral arms for example) and are hard to quantify.
The uncertainties of the 24~$\mu$m flux, as given by SExtractor, are available in the on-line Tables 
of \citet{2011A&A...534A..96S}. As described by \citet{2011A&A...534A..96S}, we apply an average correction to stellar masses of low-luminosity YSCC for the IMF incompleteness.
While uncertainties on the distribution of cluster masses and on the mass of bright individual YSCs are of order 0.1~dex, the IMF incompleteness implies larger uncertainties 
on individual YSCC masses when L$_{bol}<10^{40}$~erg~s$^{-1}$ \citep[][see their Figure ~11]{2011A&A...534A..96S}. 

As expected, YSCCs are associated with GMCs of B, C, or D type. We have only 5 YSCCs which lie at the boundary of  A-type GMCs and the cloud testers
considered  that these peaks were probably not associated with GMCs.
We expect a correspondence between b-type YSCCs and B-type GMCs, or c-type YSCCs and C-type GMCs. And that is what indeed happens, even though the cloud 
classification was not based on the correspondence with MIR-sources in the \citet{2011A&A...534A..96S} catalogue.
However, there are a few exceptions, and the apparent non-correspondence between  the cloud and the associated YSCC classification needs some explanation.  
Sometimes a cloud hosts more than one source: if a b-type  YSCC and a  c-type YSCC
are associated with one cloud then the cloud is classified as C-type. Sometimes a YSCC is identified with a MIR-source which effectively is a blend of two or 
more sources and only  one of these
has an optical counterpart. In this case the cloud and the YSCC might not belong to a similar class. 
There are also a few clouds classified as B-type or C-type with no associated YSCC. Clearly for most of these  cases the MIR emission 
at 24~$\mu$m was too weak for the emitting area to be classified as "source".  In a few cases,  blending/confusion with  
nearby sources might have caused the failure of the source extraction algorithm (see \citet{2011A&A...534A..96S}  for references and a description of the Sextractor software).
Only 18 out of 87 GMCs of B-type are not associated with YSCCs and 46 GMCs out of 286  of C-type do not host a  YSCC.  

We can summarize that 332 GMCs are associated with at least one  YSCC. 
We have 58 clouds  with a  weak non catalogued MIR-source (with H$\alpha$ 
counterpart) and 176 clouds which do not have associated emission in UV, optical or MIR and are considered inactive.

\subsection{The association between GMCs and YSCCs: the spatial correlation function}

In order to  quantify the link between GMCs and YSCCs, we analyse statistically the association by computing
the positional correlation function of the two distributions. We then compare this to what we expect for a random distribution.
This approach is fully justified provided that the distribution of each class of sources, taken separately, is spatially homogeneous.
This is not true for two reasons: ($i$) the density of objects changes with the galactocentric distance and ($ii$) GMCs  and YSCCs  are mostly 
located along the spiral arms which implies that even the clouds which have not yet formed stars are expected to lie closer to a YSCC with respect to a randomly 
distributed population in the disk. However, since the average distance between sources of these two classes is much smaller than
these large scale variations, the analysis that we are going to present is reasonably justified. In our treatment we take into account the density dependence 
from the galactocentric distance but we do not apply any correction for the spiral pattern modulation,  we  only briefly discuss it in our analysis.
In the next Section we compare with the positional correlation between GMCs and other populations in the disk, in order
to check that indeed the correlation with YSCCs selected via  MIR emission is the strongest.

To compute the expected positional correlation function of YSCCs and GMCs
in a galaxy disk, one has to consider first $f_{YSCC}(R)$, the radial density 
distribution of YSCCs. This is an azimuthal average of the deprojected local surface density i.e. of the radial surface density on the galaxy plane. One can then compute
$N(d,R)$, the average number of YSCC within a circle of radius $d$ centered on a randomly selected GMC, and $P(d,R)$, the probability of finding the closest YSCC to a GMC 
at a distance $d$. 
Having $P(d,R)$, it is then straightforward to retrieve the cumulative probability function $C(d,R)$, which is
the probability of finding a YSCC within a distance $d$ of a GMC, or equivalently the fraction of GMCs which have at least one YSCC within a distance $d$.  
We can then compare $C(d,R)$ with what is observed in a galaxy disk. 
 
Since we have a limited number of YSCCs and GMCs in the M33 disk, we
take into account the density dependence on the galactocentric distance by dividing M33 into 3 radial intervals:
1) $R<1.5$~kpc; 2) $1.5\le R < 4$~kpc; 3) $R\ge 4$~kpc;  
The number of YSCCs are  105, 290, and 236 in zones 1, 2, and 3 respectively, and in each zone we compute the 
mean density of YSCCs  as the ratio between the number of YSCCs and the disk area of the zone.
We refer to the mean YSCC density using the symbol $ \langle f_{YSCC}\rangle_i$  where $i$=1, 2, 3 for zones 1, 2, and 3 respectively and the angle brackets 
indicate that it is an average over the zone.  We define $\bar d$  as the radius of a circle inside 
which there is on average one YSCC for a randomly distributed population.  The length scale $\bar d$ is a typical separation length for YSCCs randomly distributed 
in a plane and in general it is a function of R.  For each zone we
determine $\langle\bar d\rangle_i=1/\sqrt{\pi \langle f_{YSCC}\rangle_i}$, an average value of the separation length, 
and find the following values:  146~pc,  218~pc, 374~pc for zones 1, 2, and 3 respectively.
It is useful to define this
length scale because reasoning in terms of normalized distances allows us  to better compare results obtained for the different subsamples.  

Therefore, in the absence of correlations, the average number of YSCCs within a circle of radius $d$ centered on a randomly selected GMC, $N(d,R)$,  can
be approximated in each zone as:
\begin{equation}
N_i(d) = \pi d^2 \langle f_{YSCC}\rangle_i = \Biggl({d\over\langle\bar d\rangle_i}\Biggr)^2,
\label{eq:N}
\end{equation}

The probability $P(d,R)$ of finding the closest source to a GMC at a distance $d$ and the corresponding cumulative probability $C(d,R)$ 
can be retrieved in general from $N(d,R)$ as:
\begin{equation}
P(d,R)=e^{-N(d,R)};\qquad C(d,R)=1-e^{-N(d,R)}.
\label{eq:PC}
\end{equation}

\begin{figure} 
\includegraphics[width=8 true cm]{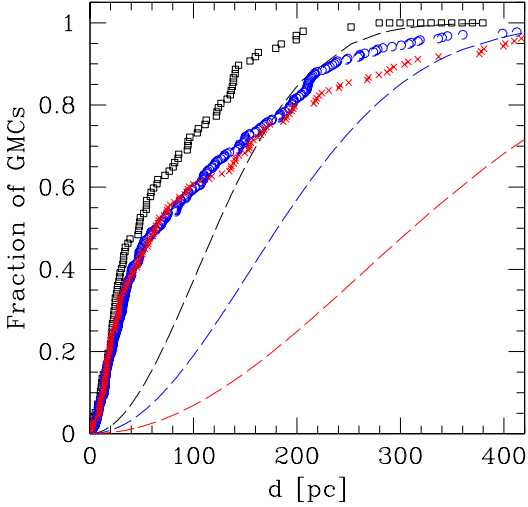}
\caption{Positional correlation of  GMCs with YSCCs. We show the fraction of GMCs with at least one  YSCC at a separation  $d$, given in pc, using black squares for 
the inner region, open blue circles 
for the middle region and red crosses for the outer region.  The dashed lines are the fractions expected for random association. A significant 
clustering is found.  }
\label{corre}
\end{figure}

In the absence of correlations, we use Eq.\ref{eq:N} and Eq.\ref{eq:PC} to compute in each zone the mean expected positional correlation functions
of GMCs and YSCCs for random distributions. These are shown as dashed lines
in Figure \ref{corre} as a function of the separation between a GMC and a YSCC given in parsecs. In the same Figure we plot the observed
fraction of GMCs with at least one  YSCC at 
a separation  $d$  using black squares for the inner region, open blue circles for the middle region and red crosses for the outer region. 
Assuming that a YSCC and a GMC are closely associated if they  are separated by no more than  a  cloud radius, which is typically $r_{e}\simeq$ 50~pc,  
we expect to find a fraction of GMCs of order $r^2_{e}/\langle\bar d\rangle_i^2$ which have a YSCC closer than 50~pc if YSCCs are randomly distributed. 
On average  we should find 12\%, 5\% and 2\% of GMCs with a YSCC within  $r_{e}$ for zones 1, 2, and 3 respectively.
Instead we observe GMC fractions of   53\%,   45\%, and  43\% for zone 1, 2 and 3 respectively. This implies that we have about 4 times more YSCCs in zone 1, 
9 times more YSCCs in zone 2, and 21 times more YSCCs in zone 3 that are in close association with GMCs than what might be expected if GMCs and YSCCs were 
randomly distributed. 
 
\begin{figure}
\includegraphics[width=10 cm]{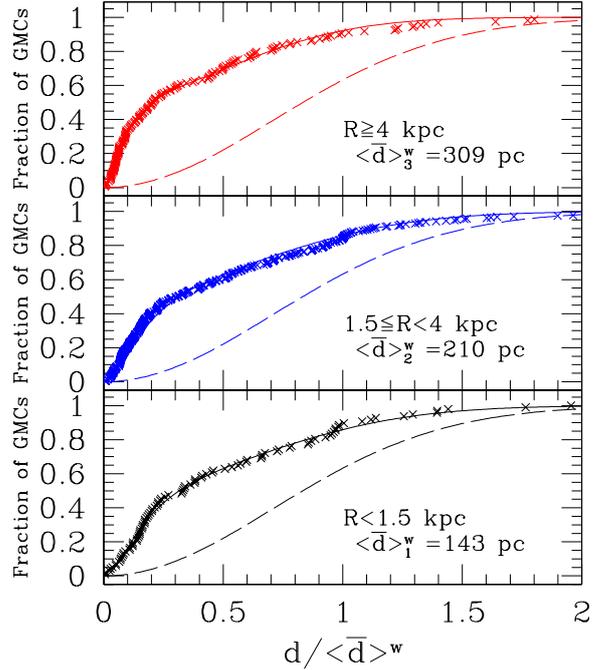}
\caption{Positional correlation of GMCs with YSCCs, for the 3 selected radial ranges in the M33 disk, as a function of $d/\langle\bar d\rangle^w$. The innermost and central 
regions have quite similar
positional correlation function if distances are scaled with $\langle\bar d\rangle^w$. The continuous lines  are the fits with our model accounting for a correlation, 
while the dashed curves represent 
the expectation in the absence of correlation.}
\label{fig:fitresults}
\end{figure}

One can also display the distributions of Eq.\ref{eq:PC} as a function of the normalized distance $d/\langle\bar d\rangle$, 
and in this case only one curve is necessary to pin down
the randomly distributed population, independently of the disk zone. But before doing this
we would like to introduce some weighted average quantities to better take into account radial variations of GMC densities,  as one goes 
from the crowded central areas of M33 to the disk outskirts. If YSCCs and GMCs are randomly but non-uniformly distributed in the disk, we can compute the mean separations 
and densities of YSCCs as seen by the GMC population.   
In this case a good approximation to the density of YSCCs in each zone is a weighted mean,  with the weights given by the GMC number densities. We call this weighted
mean $ \langle f_{YSCC}\rangle^w_i$ (with $i$=1, 2, 3 for zones 1, 2, and 3 respectively). Analoguous to $f_{YSCC}(R)$, we define $f_{GMC}(R)$, the radial density distribution of GMCs
in the galactic plane. The weighted average density of YSCCs in the surrounding of GMCs for a
random distribution can then be estimated as:
 
\begin{equation}
\langle f_{YSCC}\rangle ^w_i=\frac{\int_{R_{min,i}}^{R_{max,i}}{R\,f_{GMC}(R) f_{YSCC}(R)\,dR}}{\int_{R_{min,i}}^{R_{max,i}}{R\,f_{GMC}(R)\,dR}}
\label{eq:wei}
\end{equation}
 
where $R_{min,i}$ and $R_{max,i}$ are the radial boundaries of each disk zone. We can then
estimate the mean separation between YSCCs and GMCs for a random distribution as $\langle \bar d\rangle^w_i=1/\sqrt{\pi \langle f_{YSCC}\rangle^w_i}$ and use this to normalize $d$. 
We have $\langle \bar d\rangle^w_1$=145~pc, $\langle \bar d\rangle^w_2$=210~pc $\langle \bar d\rangle^w_3$=309~pc.  
Thus, the difference with the earlier calculations is small and mostly relevant for zone 3.
Using the weighted average YSCC density for each zone, as in Eq.\ref{eq:wei},  the quantities $N(d)$, $P(d)$ and $C(d)$ can be computed from Eq.~\ref{eq:N} and \ref{eq:PC}. 
In Figure \ref{fig:fitresults} the random distribution is shown as a function of
$d/\langle\bar d\rangle^w$. The fraction of GMCs which have at least one YSCC at a distance $\langle\bar d\rangle^w$ for a random distribution is 0.63, 
and almost all YSCCs are at a distance $d< 2 \langle\bar d\rangle^w$ from a GMC. The crosses in Figure  \ref{fig:fitresults}
show the observed cumulative functions in the three disk zones. 
As already stated, the true distributions are far  from random since about half  of the YSCCs are within 0.25 $d/\langle\bar d\rangle^w$ of a GMC.

In what follows we  model the observed positional correlation function of Figure~\ref{fig:fitresults} and determine the correlation length. 
For a simplified approach, we propose the following form for the observed average density of YSCCs at a distance $d$ from a randomly chosen GMC:
 
\begin{equation}
F(d)=\frac{c_0}{\strut\pi\bar d^2}+\frac{c_1}{2\pi\lmbc^2}e^{-d/\lmbc};
\end{equation}
 
where $\lambda_c$ plays the role of a correlation length, and the length scale $\bar d$  is, as stated before, the typical separation length for YSCCs randomly 
distributed in a plane.
In the absence of correlation, we should expect $c_0=1$ and $c_1=0$; in the case of a positive correlation, instead, we expect to have $c_1>0$, as well as $c_0<1$ 
to balance the average density on randomly chosen positions. The quantity $N(d)$  can be computed by integration, giving:
 
\begin{equation}
N(d)=\int_0^d{2\pi d' F(d')\  {\hbox{d}} d'} = c_0{d^2\over \bar d^2}+c_1\left(1-(1+{d\over \lmbc})e^{(-d/\lmbc)}\right).
\label{eq:nd}
\end{equation}
 
Using the values of the weighted mean YSCC separation $\langle \bar d\rangle^w_i$
in Eq.~\ref{eq:nd}, we derive as usual the expected $P(d)$ and $C(d)$  for each zone. By comparing the modelled $C(d)$ to the observed fractions of GMCs which have  
 at least  one YSCC within a given value of  $d/\langle \bar d\rangle^w$, we retrieve the average values of $c_0$, $c_1$ and $\lambda_c$ in each zone.  
A least square fitting method of the data is used to  determine $c_0$, $c_1$ and $\lambda_c$ in the modelled $C(d)$.  
 
The values of $\langle \bar d\rangle^w$
and  all parameters derived from our fits for the three zones are listed in Table~\ref{tbl:fitresults}. The continuous lines in Figure~\ref{fig:fitresults}  shows the very good 
quality of the fits and, 
for comparison,  also the expectation in the absence of correlation (dashed line). It is easy to see that introducing a correlation one obtains fits far better than without it.
The correlation length comes out to be very similar in the three zones: 15.8, 17.7, and 17.9~pc, which seems to outline the physical relation between the GMC and YSCC within the same 
star-forming region. There is a highly statistically significant clustering of  GMCs and YSCCs at larger distances. 
The relative density contrast of the correlated pairs, within a circle of radius $3\lmbc$ (around 50\,pc in all three cases) is 82.3\%, 87.7\%, and 91.4\% respectively.
The reason why  $c_0$ is larger than unity in all cases, while it should have been lower than that, is likely an effect of radial density variations within each zone.
The mean separations  vary across the disk and have been averaged combining regions with rather different local source densities.  This is especially true at large galactocentric 
distances and it is enhanced by filamentary overdensities, justifying why the largest value of $c_0$ is found for the outer region (here the radial density of sources
drops and density variations are higher).

The steeper increase of the positional correlation function around 100-200~pc in the three zones is likely due to the presence of spiral arms or gas rich filaments.
These filaments host many CO clouds and YSCCs and have thicknesses of 100-200~pc.   
The positional coincidence of YSCCs and GMCs in the outer disk is extraordinary and chance alignments much less likely than in the inner disk.
It would be interesting to analyze the effects of both GMC and YSCC crowding in the spiral arms. The present treatment is however not suited to investigate quantitatively 
such effects, because the correlation is taken to be isotropic  while in the presence of a spiral pattern this assumption should be released. 
We plan to devise a 2-dimensional extension of the present analysis in a future work, aimed at outlining also the correlation pattern at intermediate distances (of order 100-200 pc). 
The short-range correlation presented here is however not appreciably affected by spiral arm crowding, since it corresponds to distances much smaller than the typical width of the arms. 
 
\begin{table}
\caption{Fits to GMCs and YSCCs positional correlation functions} 
\centering                                       
\begin{tabular}{c c c c c }           
\hline\hline 
zone &   $\langle \bar d\rangle^w$ [pc] & $c_0$  &  $c_1$ &  $\lmbc$[pc] \\                  
\hline\hline 
$R<1.5$\,kpc          & 143    &  1.23    &  0.73    & 15.8  \\
$1.5\leq R<4.0$\,kpc  & 210    &  1.17    &  0.67    &  17.7   \\
$R\geq4.0$\,kpc       & 309    &  1.82    &  0.73    &  17.9   \\
\hline\hline 
\end{tabular}
\label{tbl:fitresults}
\end{table}

If we subdivide the clouds into a high-mass sample and a low-mass sample, according to whether the GMC has a luminous mass higher  
or lower than 2 $10^5$~M$_\odot$, we find that 69$\%$ of the high-mass GMCs have a YSCC within 50~pc, while this happens only for 44$\%$ of the low-mass 
GMCs. This implies that it is rarer to find inactive GMCs of high mass than of low mass. For random association the percentage is much lower, of order 3$\%$.

\section {The association between GMCs and other types of sources in the disk}

In nearby galaxies the spatial resolution of today observations is high enough to identify  the various products of the gas-star formation cycle, such as
H$\alpha$ regions, massive stars, embedded  or optically visible stellar clusters. \citet{2003ApJS..149..343E} used the BIMA disk survey of the inner 
disk of M33 to correlate the position of 148 GMCs with  HII regions identified through H$\alpha$ emission. They found a significant clustering especially 
for the high mass GMCs. Given the high number of identified HII regions (about 3000 in the innermost 4~kpc), the difference in the number of HII regions closer
than 50~pc to a GMCs with respect to the random distribution is only a factor 2. Hence the clustering of our sample of YSCCs  around GMCs is much stronger than that of
HII regions and this can be easily understood since the YSCCs have been identified through MIR emission, present in the early phases of star-formation,
while less compact H$\alpha$ sources, like shells and filaments, are formed at a later stage during the gas dispersal phase.  

\citet{2012ApJ...761...37M} analyzed the location of 65 massive GMCs in the inner disk of M33 with respect to massive stars which they identify through the optical surveys of 
\citet{2006AJ....131.2478M} in various optical bands. The aim is to study the evolution and lifetime of GMCs by estimating the ages of the closest bright stars using
stellar evolution models. In particular they confirm a scenario of recursive star formation since dense molecular gas, the fuel for the next stellar generation,
is found around previously generated massive stars. The \citet{2012ApJ...761...37M} clusters, with estimated stellar masses in the range $10^{3.5}-10^{4.7}$~M$_\odot$ and ages
between 4 and 31~Myrs, are on average older and more massive than YSCCs associated with MIR sources. Unfortunately the authors do not list the positions of OB
associations and clusters and do not determine the  statistical significance of the clustering around GMCs.

\begin{figure} 
\includegraphics[width=9cm]{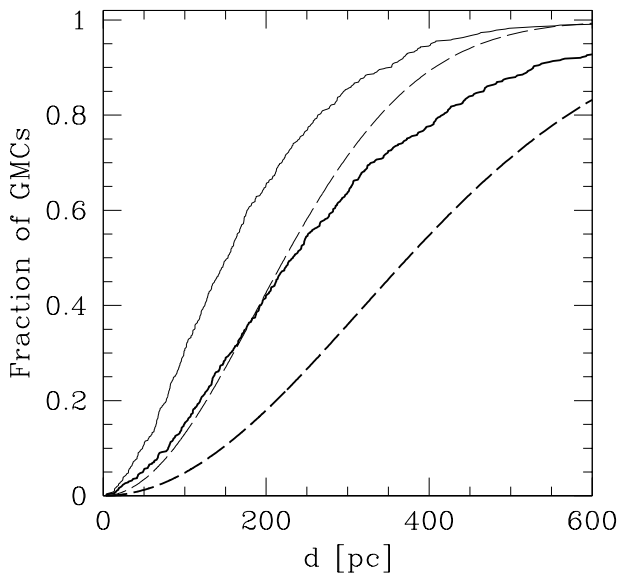}
\caption{Positional correlation of  GMCs with optically visible stellar clusters. We show the fraction of GMCs with at least one YSC at distance $d$ shown as a  
function of $d$ in parsecs. The dashed line is the fraction expected for random association.  The thicker lines refers to YSC with ages less than 100~Myrs.}
\label{yscass}
\end{figure}

To check whether optically visible stellar clusters lie  near GMCs, we use the compilation of \citet{2014ApJS..211...22F}  who identified 707
stellar cluster candidates. They determined ages and masses of 671 of them through  UBVRI photometry using archival images of the Local Group Galaxies 
Survey \citep{2006AJ....131.2478M}. Some of these clusters have ages $\le 100$~Myrs. We find
that 668 of the 707 clusters lie in the area of the CO survey and of these only 64, about 10$\%$, are separated by less than one cloud radius 
from the nearest GMC. The estimated ages of these 64 clusters vary between 5~Myrs and 10~Gyrs with a mean value 
of about 50~Myrs. Since the  expected fraction of GMCs  within 50~pc distance from a stellar cluster  is only
a factor 3 smaller for a random distribution, the statistical significance of the association between GMCs and optically visible clusters is far less than with 
YSCCs associated with MIR sources.  We also check the association considering only optically visible stellar clusters 
whose ages are less than 100~Myrs (about 1/3 of the sample), as displayed by  the thicker lines in Figure~\ref{yscass}. The ratio between the observed fraction of GMCs  
in the proximity of optically selected YSCs, and the expected fraction for a random distribution, is larger than for the whole sample of optically selected clusters. 
Ever in this case, however, optically visible clusters are less correlated with GMCs than infrared selected YSCCs examined in 
this paper. Moreover, the peak of the excess of the observed fraction of GMCs, with respect to a random distribution, is at distances of order 200-300~pc, much larger 
than a typical cloud radius, as shown in Figure~\ref{yscass}. The weak correlation found is then likely driven by the location of GMCs and YSCs along gaseous
filaments or spiral arms rather then by a one to one correspondence.

\section{Cloud classes and their properties}

In this Section we examine cloud classes as being representative of different stages of cloud evolution and their properties, such as their location across 
the disk and the associated luminosities.
We have classified the population of molecular clouds in M33 into 3 categories according to whether they are non-active,  A-type,
have MIR emission without optical counterpart, B-type, or have some MIR emission with associated H$\alpha$ emission, C-type. These 3 classes
may correspond to three different stages of molecular cloud evolution. 

The 172 clouds classified as A-type are considered inactive because they may 
have   just formed but have not yet fragmented to form stars. 
In this category we might have  some clouds which are left over from a previous
episode of star formation; the stars  break out from the cloud and the most massive  stars could 
enhance the formation of molecular material close to bright H$\alpha$ filaments by compressing  the ISM through  winds and  expanding bubbles. 
Some radiation from nearby stellar associations might heat the dust triggering some diffuse 
MIR emission. We associate the 87 clouds of B-type to the early phases of star-formation, where radiation of
massive stars is not yet visible in optical or UV bands because it is   absorbed by the dust of the surrounding cloud material. In this case the
radiation  heats the dust in the clouds which emit in the MIR in localized areas. In the 286 clouds of class C, ultraviolet emission 
from young stars or H$\alpha$ line emission from ionized gas is visible within the cloud contours. 
In this case, winds from evolved stars can  sweep out  gas in their vicinity and UV light   escapes  from 
the cloud. The H$\alpha$ radiation, less absorbed by dust than UV continuum, may become detectable  when the HII region is still compact. 
Winds from young massive stars can ultimately disrupt the parent cloud and quench star formation.
If the stellar cluster is of small mass,  massive stars are not always present 
\citep{2009A&A...495..479C}.  In this case, the 
lack of H$\alpha$ radiation might push the classification of evolved clouds into class B and cause the MIR emission to be weak.
Moreover, since clouds are not spherically symmetric and stellar clusters are  not necessarily born at cloud centers, geometrical effects may play a role 
in mixing somewhat the B and C type.  On the other hand, emission on the line of sight to the cloud but unrelated to the cloud (i.e. foreground or background) could result
in an overestimate of the state of evolution of the cloud. This implies that  in each class there will be a few clouds whose evolutionary stage might be inappropriate.
 
\begin{figure} 
\includegraphics[width=9cm]{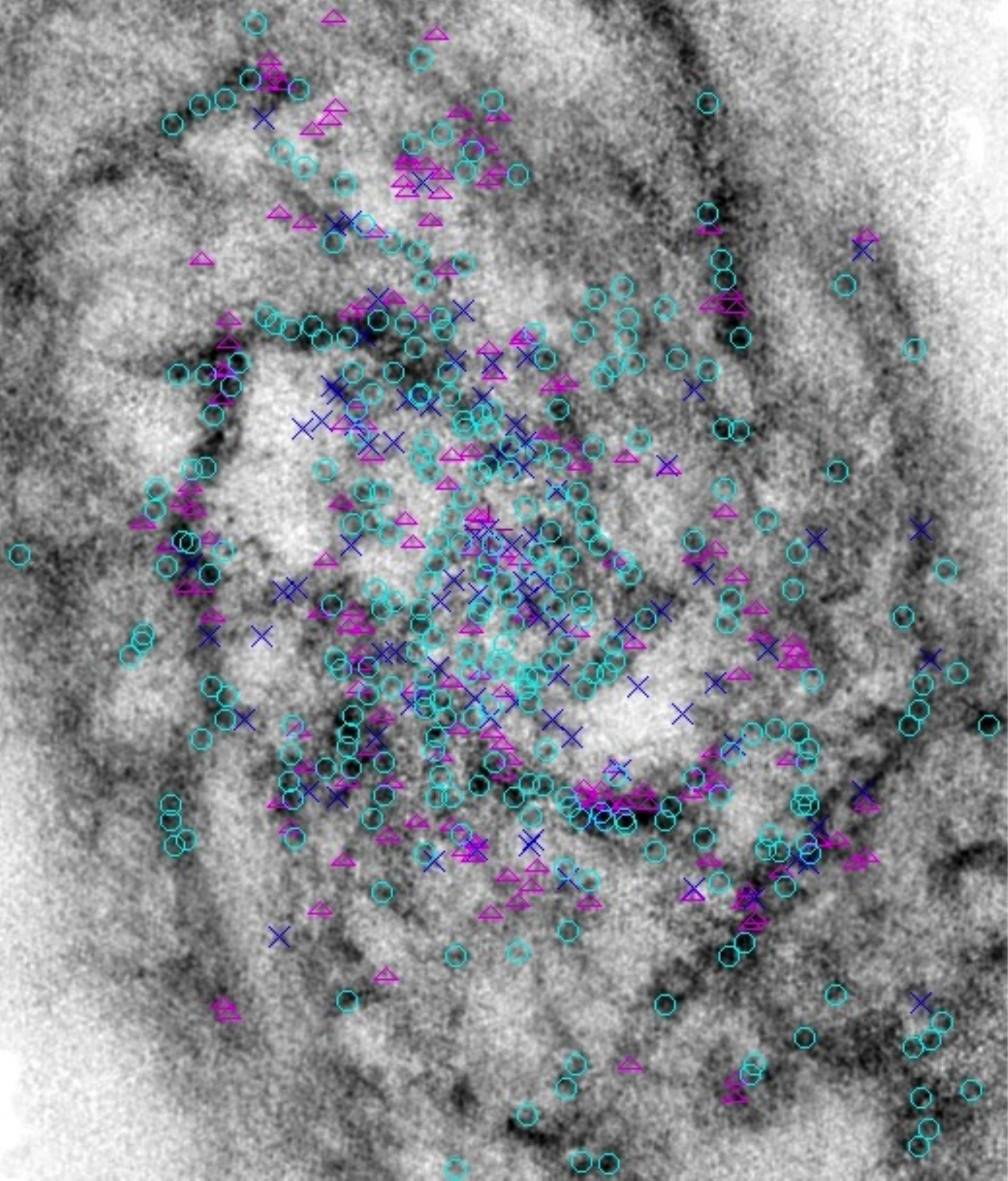}
\caption{The location of the three cloud classes across the M33 disk shown over the HI image of the galaxy.
Open cyan circles are C-type clouds, blue crosses are B-type clouds, magenta triangles are A-type clouds.}
\label{cltype} 
\end{figure}

\begin{figure} 
\includegraphics[width=9cm]{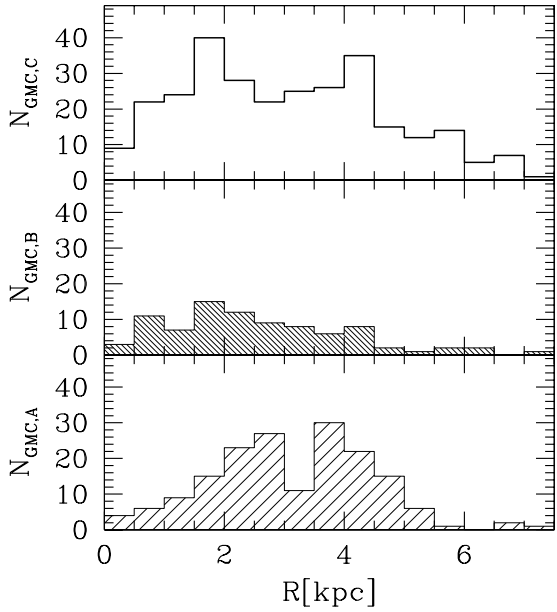}
\caption{Number of GMCs of A-type, B-type and  C-type in 0.5~kpc bins of galactocentric radius are shown in the bottom, middle and upper panel respectively. }
\label{clbin} 
\end{figure}

There are two evident differences in the location of the 3 types of clouds across the M33 disk. The GMCs are plotted in Figure~\ref{cltype} over the HI image of M33
and in Figure~\ref{clbin} as a function of galactocentric radius. 
The A-type clouds have a median galactocentric distance over 3~kpc with only 20\% within 2~kpc of the center, where the B- and C-type clouds are more numerous (see Fig.~\ref{clbin}). 
Both the A- and C-type clouds are found along HI filaments and along the northern and southern spiral arm
of this flocculent galaxy. Along the southern arm, which is more regular and  less disturbed than its northern counterpart, the A-type clouds are found in the
inner part of the arm, while  C-type clouds are  more often found on the arm, where the peak of the HI and H$\alpha$ emission is. This is in agreement with
the theories of formation of molecular complexes across the spiral arms \citep{1969ApJ...158..123R,2002ARA&A..40...27C}: the atomic gas experiences a first 
compression in the inner part of the arm 
and becomes molecular, then as it enters the arm  the supersonic turbulence enhances the fragmentation and the process of star formation.  
This scenario is not seen for the less regular northern arm.
A-type clouds are also more clustered than C-type clouds which are more coarsely placed along the gaseous filaments.  
Clouds of high  mass are very rare at large galactocentric radii,  possibly because there are no spiral arms that trigger their growth.  
 
The B-type clouds are rarely found close to the spiral arms   or the brightest 21-cm peaks but they are more numerous along filaments of moderate 
HI surface density. They  are present at all radii smaller than 4~kpc and are never a dominant population. If they represent a transition phase between inactive
and active clouds the lack of these clouds along the arm means that this transition must be a   rapid one. 
Lower mass stellar clusters, not bright in the UV or H$\alpha$, might be  associated with these clouds, but the star formation
process may continue to increase the YSCC mass at a later stage.  
Although there might be some  contamination among MIR-sources with no optical counterpart, due to obscured distant galaxies in the background of M33,
the presence of CO emission at the same location of the MIR-sources assures that most of these  are  YSCCs associated
to M33. Their clustering along HI overdensities underlines again their association with the M33 ISM.

We now examine the CO-luminosities of GMCs in the three classes as well as their emission at 24~$\mu$m. The mean  GMC mass is progressively
increasing from 1.3$\times 10^5$~M$_\odot$ for A-type clouds to 2.1$\times 10^5$~K~km~s$^{-1}$~M$_\odot$ for B-type to 3.6$\times 10^5$~~M$_\odot$
for C-type clouds. Hence, B-type clouds are   typically of intermediate mass, and not necessarily of low mass.
If we interpret the sequence of clouds from A-type, to B-type, to C-type, as a time sequence, the increase of the average cloud mass going from
A-type to B- and to C-type implies that GMCs collect more gas as they age. The trend   remains if we look only at inner disk clouds (at $R<3.5~$ kpc) 
or outer disk clouds (at $R\ge 3.5~$ kpc).
 
\begin{figure} 
\includegraphics[width=10.8 cm]{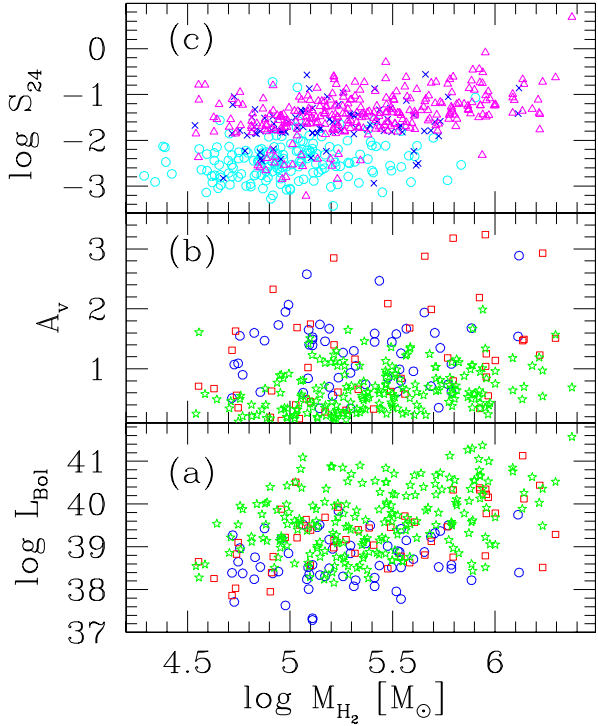}
\caption{The YSCC bolometric luminosity, in erg~s$^{-1}$, and visual extinction, are plotted in panel $(a)$ and $(b)$ respectively as a function of the 
total luminous mass of GMC hosting the YSCC.  
Blue circles are embedded b-type sources; red squares are cluster at birth c1-type sources; green stars are young clusters c2-type sources.
In panel $(c)$ the flux density at 24~$\mu$m in units of mJy~arcsec$^{-2}$ (see text for more details) is plotted as a function of the total GMC luminous mass.  
Open cyan circles are C-type clouds, blue crosses are B-type clouds, magenta triangles are A-type clouds. }
\label{comir} 
\end{figure}

Clouds of C-type extend to higher molecular masses than B-type clouds and Figure \ref{comir}  shows the total luminous masses of the clouds versus
the YSCC bolometric luminosity (for GMCs hosting these),  extinction  and flux density at 24~$\mu$m. We compute the MIR flux densities of YSCC associated with B
or C-type sources by dividing the flux as measured by \citet{2011A&A...534A..96S} by the source area (defined at 8-sigma isophote). 
The A-type clouds have very low MIR flux densities except a few which are contaminated by nearby sources. 
For these and for clouds with no associated YSCC we  measure  the emission at 24~$\mu$m in
apertures of 3, 5, and 8 pixels (1~px=1.6~arcec) in radius centered on the cloud. 
After background subtraction, we choose the aperture
which gives the largest flux density; this is derived by dividing the flux in the aperture by the aperture area. 
With this procedure we underestimate the  MIR flux density of the emitting area because we cannot measure the effective source size (the emission is
indeed so weak that a source size is hard to define and in fact flux levels are below that of sources found by the \citet{2011A&A...534A..96S} procedure).
The GMCs associated with catalogued sources have flux densities higher than 0.01 mJy/arcsec$^{-2}$ (425 MJy/sr).  
This flux limit for catalogued sources, which is evident 
in panel $(c)$ of Figure~\ref{comir}, is a result of the source extraction algorithm \citep{2011A&A...534A..96S}. GMCs with no catalogued sources are 
mostly of A-type but there are also a few C-type clouds with weak MIR emission, perhaps being in the process of dissolving due to the  evolution of the 
newborn cluster.  A correlation between the cloud CO luminous  mass and the source extinction is found for sources of c2-type, as shown in Figure~\ref{comir}.
The same figure also shows there is no correlation between the GMC mass and the YSCC bolometric luminosity. Similarly,  there is no correspondence 
between GMC masses and YSCC masses when, through successful SED fits, stellar masses  have been determined \citep{2011A&A...534A..96S}.
This implies variations of the amount of gas turned into stars within giant complexes where star formation is not uniformly spread out.  

There is no difference in the CO line-width for clouds of class A, B, C (3.5$\pm1.5$ 3.3$\pm1.1$ and 3.4$\pm$1.2 km~s$^{-1}$  respectively). The 
slightly larger dispersion for clouds of A-type  is due to a secondary peak in the width distribution at narrow linewidths ($\sigma_v^{gau} \sim 2$~km~s$^{-1}$). 
Similarly there is not much difference in  size between  complexes of different types except that GMCs of large size are of C-type. Sizes are distributed 
between 10 and 100~pc with a peak at 30~pc and a mean value of order 40~pc for  A- and B-type clouds, and of 50~pc for C-type clouds.

\section{YSCC classes and their properties}

In Section 3.1 we defined classes of YSCCs.  YSCCs of  b-, c1-, c2- and d-type are associated with GMCs, while those of  c3- or
e-type are not. The YSCCs of b- and e-type do not have an optical counterpart or only very faint emission in the SDSS  images. 
YSCCs of  c-type are associated with blue light in SDSS images and have H$\alpha$ and FUV emission.  Sources of d-type are  
ambiguous between b- and c-type. Since sources of  b-type are associated with GMCs, they  presumably 
represent the early stages of SF, when stars are fully embedded in the cloud.  e-Type YSCCs may be proto-clusters in their embedded phase
associated with lower mass molecular clouds, which have not been detected by the IRAM-survey. 
Like b-type sources, the e-sources tend to lie along HI overdensities but not along the brightest ones. The MIR-sources identified as e-type 
YSCCs often lie  near a SF region and it is not clear whether they host protostars or if the dust is heated by    
young stars in their proximity. A few are rather isolated and may be background obscured objects. They are more numerous than b-type YSCCs 
in regions far from the center.  The MIR emission of  e-type YSCCs is on average lower than that of  b-type YSCCs, and hence they might be associated 
with small clouds \citep{2011A&A...528A.116C}, with a lower metallicity \citep{2010A&A...512A..63M} and dust content and undetectable through the IRAM-CO survey.   

\begin{figure} 
\includegraphics[width=9 cm]{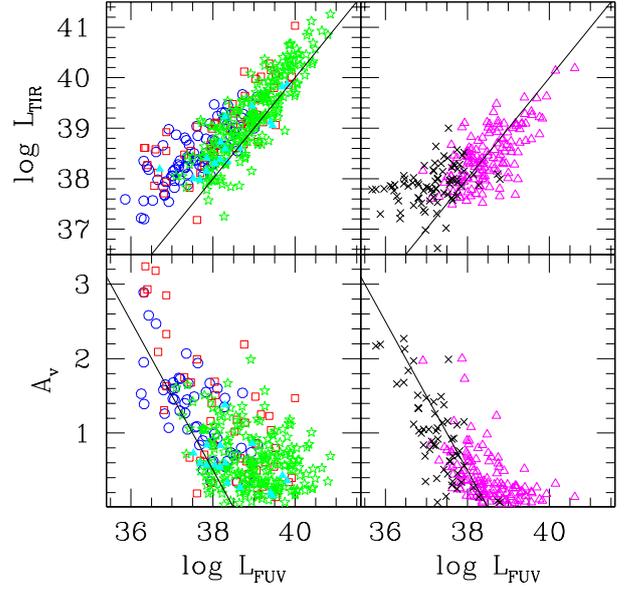}
\caption{YSCC visual extinction and Total InfraRed luminosity as a function of FUV luminosity.
In the lower and upper left panels we show sources associated with GMCs, in the lower and upper right panel we show sources
not associated with GMCs. Blue circles: embedded b-type sources; red squares: cluster at birth c1-type sources; green stars: young clusters c2-type sources;
cyan filled triangles: unclassifiable sources with GMCs; magenta triangles: clusters without GMCs; black crosses: sources of unknown nature without GMCs (e-type).
Continuous lines in all four panels are not fits to the data but are for reference having unity slopes.   }
\label{opa2}
\end{figure}

\begin{figure} 
\includegraphics[width=9 cm]{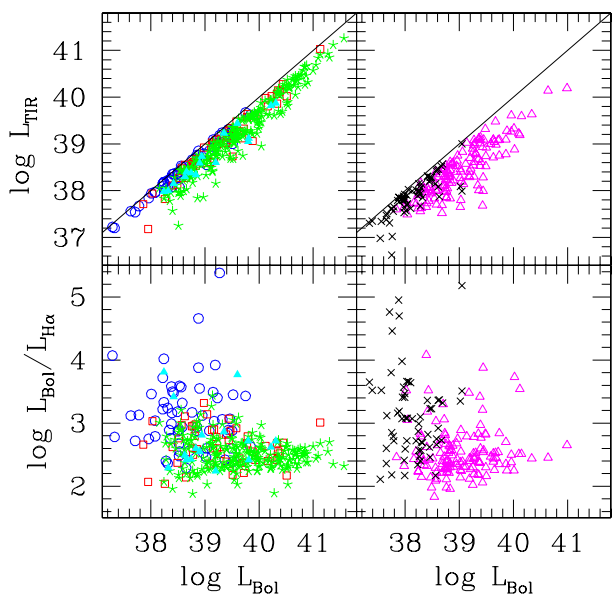}
\caption{YSCC Total InfraRed luminosity and Bolometric-to-H$\alpha$ luminosity ratio are plotted as a function of the MIR-source bolometric luminosity.
In the lower and upper left panels we show sources associated with GMCs, in the lower and upper right panel we show sources
not associated with GMCs. Blue circles are embedded b-type sources; red squares are cluster at birth c1-type sources; green stars are young cluster c2-type sources;
cyan filled triangles are unclassifiable sources with GMCs; magenta triangles are clusters without GMCs; black crosses are sources of unknown nature without GMCs (e-type). 
Continuous lines in all four panels are not fits to the data but are for reference having unity slopes. }
\label{opa3}
\end{figure}

If the sequence  b, c1, c2, c3, represents a time sequence in the evolution of a star forming region, we should find progressively less extinction
and older ages for the associated clusters.
In Figure \ref{opa2} we show the YSCC FUV luminosity versus extinction A$_v$, as given by Eq.(8) of \citet{2011A&A...534A..96S}, similar to the visual 
extinction definition used by \citet{2001PASP..113.1449C}.
In Figures \ref{opa2} and \ref{opa3} we can see that YSCCs of b and c1-type are the most embedded, having the highest values
of A$_v$. There are, however, two differences between these types of sources: some YSCCs of c1-type have   
high bolometric luminosities and hence may have already formed massive stars. On the birthline diagram \citep{2009A&A...495..479C}
the most luminous c1-type YSCCs occupy
the same area as c2-type YSCCs while the low luminosity ones tend to overlap with the region where b-type sources are. This  implies  that c1-type YSCCs
are intermediate between fully embedded clusters in the process of formation (b-type) and very YSC (c-type). 
It is likely that c1-type YSCCs do not have FUV emission  at the location of the MIR peak because of extinction.  The YSCCs of c2-type have A$_v$ values
always lower than 2 and often lower than 1, with an average value of 0.62$\pm$0.38. Their infrared luminosity correlates very well with the FUV luminosity.
There are no YSCCs with very low Total Infrared (hereafter TIR) or FUV luminosity in this class   \citep[see][for the TIR luminosity computation]{2011A&A...534A..96S}. 
The lack or faintness of H$\alpha$ emission for b-type sources is coherent with
the picture that in these sites massive stars have not formed yet.

\begin{figure} 
\includegraphics[width=9 cm]{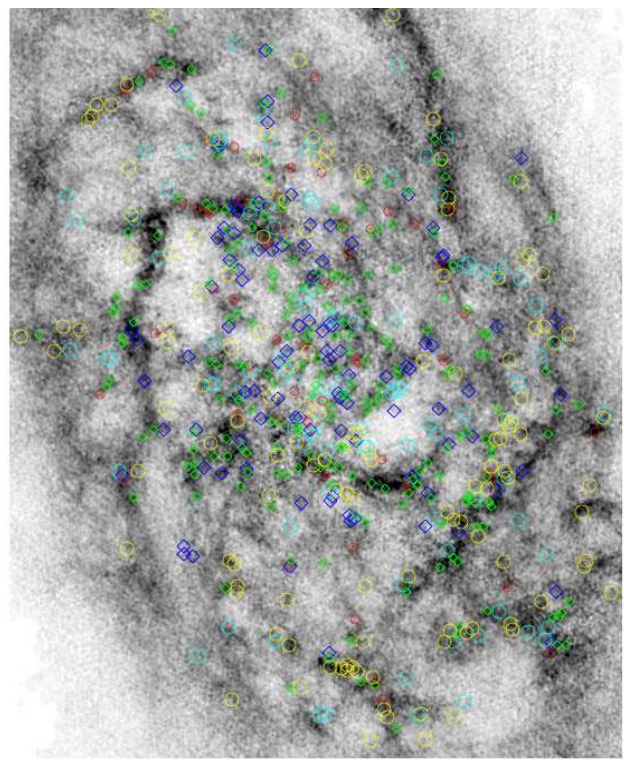}
\caption{The YSCCs in the area covered by the IRAM-all disk survey color coded and with symbols according to their class type, are  plotted over the HI map. 
Romb symbols are for YSCCs associated with GMCs
(blue for b-type, red for c1-type, green for c2-type). Circles are for YSCCs not associated with GMCs (yellow for c3-type, cyan for e-type).}
\label{src_hi} 
\end{figure}

The YSCCs not associated with GMCs are of c3-type, star forming with optical counterpart,  or of e-type, without optical counterpart.
The YSCCs of c3-type mostly have low extinction, and this is consistent with the fact that they are not associated with GMCs. They are not 
of low luminosity and hence they might be slightly more evolved than c2-type sources or associated to low mass clouds highly efficient in making stars.
In Figure \ref{src_hi} we plot the location of all YSCCs, color coded according to 
their type, superimposed on the HI map of M33. There are some differences between the various YSCC classes:  c1- or c2-type YSCCs lie over bright inner 
HI filaments or arms. The b-type YSCCs
populate  the inner disk and there are very few beyond 4.5~kpc, but they avoid the brightest HI ridges. The c3- and e-type YSCCs  are the dominant 
population at large galactocentric radii. It is  likely  that at these radii  molecular clouds are of
low mass to escape detection in the IRAM all-disk survey.  e-Type YSCCs have more diffuse MIR emission, often mixed with  faint H$\alpha$
filaments and hence it is unclear if at the location of these sources any star formation will ever take place. 

\begin{figure} 
\includegraphics[width=9 cm]{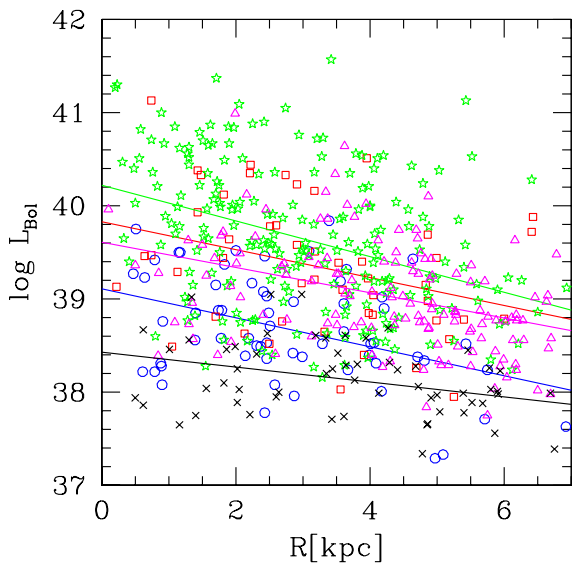}
\caption{ YSCC bolometric luminosities in erg~s$^{-1}$ as a function of galactocentric radius color coded according to the various classes as in Figure~\ref{opa3}. 
Straight lines show the linear fits to the displayed distribution for each class.}
\label{bolrad}
\end{figure}

We now examine the luminosities of the YSCCs for the various classes. In Figure \ref{bolrad} we plot the bolometric luminosity \citep{2011A&A...534A..96S}
as a function of galactocentric
radius for the various  classes and the linear fit to the log distributions.  We have included all YSCCs with an estimated bolometric luminosity which 
lie at galactocentric distances R$<7$~kpc. We have not considered the uncertainties on the bolometric luminosities   because they are hard to quantify as 
they are not dominated by photometric errors but by uncertainties related to the modelling.
On average and at all galactocentric radii the class with the highest luminosities is the c2 class, followed by c1, c3, b, and e classes. Since YSCCs of c-type 
are likely to be associated with
clusters which have completed the formation process, it is conceivable that they are more luminous than b- and e-type sources linked to the embedded phase of star
formation, when not all cluster members have been formed. In particular this finding is in agreement with non-instantaneous cluster formation theories which predict
that massive stars are fully assembled  at a later stage and their quick evolution switches off the cluster formation process \citep{2007ARA&A..45..481Z}. 
A similar trend to that shown in Figure \ref{bolrad} is found if we plot the TIR luminosity.  

\begin{figure} 
\includegraphics[width=9 cm]{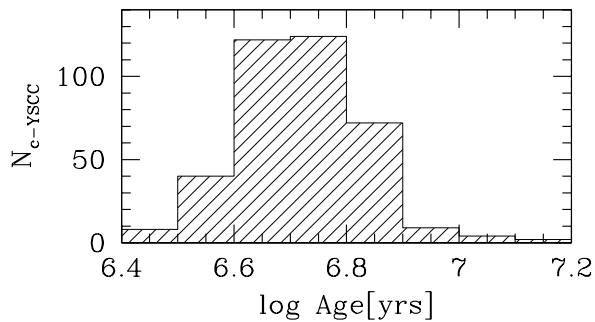}
\caption{Age distribution of YSCCs of c-type (c1,c2,c3). Only 15 YSCCs, less than 4$\%$, are older than 7~Myrs.}
\label{age}
\end{figure}

Through SED fits, \citet{2011A&A...534A..96S} have estimated the ages of YSCCs  of c-type which are shown in Figure \ref{age}. Given
the uncertainties in  age determination (of order of 0.1 dex for bright rich YSCCs and  larger for dim  ones due to IMF incompleteness) the mean ages of YSCCs 
associated with c1,c2 or c3-type YSCCs are consistent. 
It is important to underline that  the vast majority  of the YSCCs have ages between 3.5 and 8~Myrs and only
15 of them, less than 4$\%$,  are older than 8~Myrs. 

\section{Molecular cloud lifetime and the  star-formation cycle}

Stars form from gas and leave their imprint on the gas through cloud dispersal or gas compression for the next generation of stars. 
During the formation phase and for the following few Myrs, the dust associated with the molecular  material is heated by the energetic photons emitted by the YSC.
The CO survey of the LMC at a resolution of 40~pc demonstrated the good correlation of  molecular clouds with young stellar clusters of age $\le 10$~Myr 
\citep{1999PASJ...51..745F}. The timescale of each evolutionary stage has been then estimated by subdividing GMCs into different classes corresponding to different ages 
of the associated YSC \citep{2001PASJ...53..985Y,2009ApJS..184....1K}. The LMC is an irregular metal poor galaxy, interacting with the Milky Way, 
and hence it is not clear that the same timescales for GMC evolution, as for M33 or the Milky Way, apply. 

One difficulty we have to face for M33 is that despite the excellent spatial resolution of the CO survey, giving a comparable physical scale to the LMC CO survey, 
there is no high resolution radio continuum database which can be used to locate thermal and non-thermal
sources, to obtain information on the very early phases of  stellar cluster formation. Dedicated optical surveys of the LMC have produced a
catalogue of clusters and OB associations with 137 of them being  of age < 10~Myrs. In M33 there are only  16 optically identified clusters with age < 10 Myrs 
and none of these are within the cloud contour of identified GMCs. The MIR-source catalogue of \citet{2011A&A...534A..96S} with numerous YSCCs is a substitute for   
the lack of optically identified YSCs in M33. However the use of these data implies that we have more information on the intermediate stages of YSC formation
but less on the dissipation phase, when the YSC emerges from the GMC and it is optically identified.
The  YSC candidates in the \citet{2011A&A...534A..96S}  catalogue have reliable age estimates, especially if they are bright and with coincident peaks in the various bands,
such as c2-type YSCCs. All 216 c2-type YSCCs have ages $\le$ 10~Myrs and 90\% between 3.5 and 8~Myrs  
and a marked peak around  5~Myrs. Similar ages are found for c3-type YSCCs, not associated with GMCs, but with an optical counterpart as well.  Figure~\ref{age} 
shows a histogram of the age distribution of c-type YSCCs.
In the previous section we have seen that the number of YSCC associated with GMCs drops when their age is $>$~8~Myrs. 
If 8~Myrs is the typical age of the cluster when 
it  breaks through the cloud,  we can say that  phase C last 8~Myrs  We hence define 8~Myrs as the timescale for a GMC of C-type.
During this stage YSC are fully assembled, including massive star formation. Shortly after this stage the YSC dissipates the associated GMCs. 
 
\begin{table*}
\caption{GMCs and YSCCs of various classes in different radial ranges} 
\centering                                       
\begin{tabular}{c c c c c c c c c c c}           
\hline\hline                         
R[kpc] &N$_{GMC}$& A-type& B-type& C-type& N$_{YSCC}$&b-type& c1-type& c2-type& c3-type& e-type\\     
\hline                                    
  $<$ 1.5&  95 &  19(20$\%$,18)   & 21(22$\%$,19) & 55(58$\%$,55)   & 98 & 21(21$\%$) & 10(10$\%$) & 39(40$\%$)  &  9(9$\%$)   & 19(19$\%$) \\        
    1.5-4&  296&  106(36$\%$,79) & 50(17$\%$,45) & 140(47$\%$,133) & 279& 53(19$\%$) & 30(11$\%$) & 115(41$\%$) &  44(16$\%$) & 37(13$\%$) \\
  $\ge$ 4&  150&  47(31$\%$,30)  & 16(11$\%$,15) & 87(58$\%$,80)   & 231& 23(10$\%$) & 15(6$\%$)  & 60(26$\%$)  &  85(37$\%$) & 48(21$\%$) \\
 \hline                                              
\end{tabular}
\label{tabrad}
\end{table*}

We can  estimate how long the inactive A-type  and the embedded B-type phases last based on the number of  clouds in the catalogue.
Considering only GMCs above our survey completeness limit, the
total number of classified clouds is 474. Of these 127 are of A-type, 79 are of B-type, 268 are of C-type. Assuming a continuous rate of star formation in M33
and that the C-type phase  lasts 8~Myrs, we estimate that the  B-type phase lasts 2.4~Myrs and  A-type phase about 3.8~Myrs.
The shortest phase for GMCs in M33 is the totally embedded phase, when the newborn cluster has
no H$\alpha$ or optical counterpart: this is a phase which was not considered explicitly by \citet{2001PASJ...53..985Y}  and  \citet{2009ApJS..184....1K}
but it is included in their inactive, Type-I phase.  We define the GMC lifetime as the time interval between the stage when most of the GMC mass has
been assembled but the cloud is still inactive, and the starting of the gas dissipation phase, when most of the molecular gas is  dispersed in the interstellar medium 
and the YSC has no large molecular clumps nearby.
This lifetime includes the cloud inactive phase, the embedded star-forming phase, and the time when the YSC breaks through the cloud and has an optical counterpart.
The cloud growth and dissipation times, which involve a substantial change of the GMC mass, have not been considered due to the GMC survey completeness limits.
We  therefore have a lifetime of about 14.2~Myrs for GMCs in M33, before they are dissipated. This is  somewhat shorter than the lifetime derived by \citet{2009ApJS..184....1K} 
 who used a sample of GMCs of masses and effective spatial resolution similar to that of our survey. 
The Type I   (similar to our A- and B-type clouds) and Type II  (C-type) evolutionary sequence of \citet{2009ApJS..184....1K}, lasts 19~Myrs.
In the Milky Way the GMC lifetime has been estimated to be in the range 10-20~Myrs \citep{1980ApJ...238..148B,1981MNRAS.194..809L,2011ApJ...729..133M}, 
shorter than earlier estimates of 30-40~Myrs \citep{1977ApJ...217..464B}.
Considering only GMCs above a limiting mass of 10$^5$~M$_\odot$ we have 115 clouds of A-type, 58 clouds of B-type, 221 of C-type. Since 
the cluster age is not a function of the associated cloud mass, assuming that C-type clouds last 8~Myrs prior to gas dispersal, we estimate that GMCs spend 4.2~Myrs in class A,
and 2.1~Myrs in class-B. This gives a total lifetime of 14.3~Myrs, very close to our previous estimate for the whole sample of GMCs above the completeness limit.
In M33 GMC lifetime prior to gas dispersal is comparable with estimates of GMC lifetimes in the Milky Way.  
 
We now subdivide the galaxy in three zones, as in Section 3: 
1) $R<1.5$~kpc; 2) $1.5\le R < 4$~kpc; 3) $R\ge 4$~kpc. 
The number of classified clouds and sources in the 3 zones is given in Table~\ref{tabrad} (unclassified objects like D-type clouds
or d-type sources are not considered here).   We give in parenthesis the percentage of the various GMC types for each  radial zone and, in
addition, the number of GMCs in that zone whose mass is above the completeness limit.
We can estimate the cloud lifetime for the three zones  considering only GMCs above the completeness limit and find that in zone 2 GMCs have the longest lifetime, 
of order 15.4~Myrs, due to the longer time clouds spend in the inactive phase. In zone 1 and 3 the GMC lifetime is of order 13.4~Myrs   and 12.5~Myrs respectively. 
In the intermediate radial range, where spiral arm are found, molecular clouds have a longer quiescent time  as more A-type clouds are found. The quiescent time is
short for GMCs in the inner regions, about half of that inferred for zone 2. In the outer regions instead, the embedded phase lasts less than elsewhere, about 1.5~Myrs. 
However, as we  will discuss in the next paragraph, the large drop in GMC
number density in zone 3 and the lower average stellar mass of the associated YSCCs increase the uncertainties in the estimated cloud type fractions and YSC ages, and
therefore in the GMC lifetime. 

Going radially outwards from zone 2 to zone 3, the number density of GMCs decreases more rapidly than the number density of YSCCs. 
This can be explained either with an unseen population of clouds of smaller mass, undetected by the IRAM-all disk survey,
or by a quicker evolution of GMCs (once stars are formed, GMCs dissolve in a shorter time as the stellar cluster evolves). 
The number of YSCC without any associated GMCs (c3-type sources), is  in fact 
the dominant population in zone 3, while in the other zones the dominant YSCC population is of c2-type.  
Let us suppose  that the larger drop in the density of GMCs at large galactocentric radii with respect to YSCC density is  due to an unseen population of  
molecular clouds, either GMCs with  weaker CO emission  or molecular clouds of
lower mass (which have not been detected by the IRAM-survey). Suppose that the percentage of missing clouds is the same for each cloud class and that the scalelength of the 
molecular gas surface density is the same at all radii (i.e. 2~kpc) and that the cloud mass spectrum  is also radially constant. In zone 3 we should find a number of clouds
equal to 75$\%$ the number of clouds of zone 2 i.e. 222 clouds. Hence there are 72 clouds which escaped detection (or even more if their mass decreases with galactocentric radius).
Of these about 42 clouds will be of C-type, 8 clouds of  B-type and 22 of  A-type. 
Consequently the number of b-type sources  would  be 31 (13$\%$), the number of c1+c2 sources  would   be 
117 (51$\%$), and the number of c3-type  would   be  41 (18$\%$) and of e-type source  would  be 40. In this case  the percentage of optically visible YSC associated with clouds  would   
be the same as in zone 2 and also the percentage of sources with or without clouds will be in closer agreement. 
So, if the hypothesis of the unseen population is correct there is not much difference in the timescale of cloud dissipation across the M33 disk.  


\section{Summary}

In this paper we present the largest database of GMCs and candidate YSCs across a galactic disk. 
Using the IRAM-30m CO J=2-1 datacube of M33 we identify 566 GMCs
and  we select 630 MIR-sources from the the \citet{2011A&A...534A..96S} list which are  young stellar clusters
in the early formation and evolutionary phases. We classified the GMCs as non-starforming (class A), with embedded SF (B), or with exposed SF (C).  
The YSCCs were put in classes based on their emission in the MIR, FUV and H$\alpha$ bands and according to the association with GMCs.
Most of the YSCC with optical and UV counterparts have estimated ages and masses.
The classification helps in drawing a possible evolutionary sequence and the relative timescales. The results of
the classification together with  the most relevant parameters of the GMCs and YSCCs  can be found in the on-line tables. 
Since M33 has a non-uniform star forming disk with varying population densities and structures going radially outwards,
we examine three distinct radial ranges: R$<1.5$~kpc, $1.5<R<4$~kpc, and $R\ge 4$~kpc  and refer to these as zone~1 (inner disk), 
zone~2 (spiral arm dominated) and zone~3 (outer disk). Below we summarize the main results discussed in the paper.  

\begin{itemize}

\item{The GMCs catalogue comprises 566 clouds with masses between 2$\times 10^4$ and 2$\times 10^6$ M$_\odot$ and radii between 10 and 100~pc.  
490 clouds are  above the survey completeness limit of  L$_{CO} \ge$ 5700 K~km~s$^{-1}$~pc$^2$ (M$_{H_2}\ge 6.3\times 10^4$~M$_\odot$). By examining the 8, 24~$\mu$m, FUV, 
and H$\alpha$ emission within 
each cloud, 545 of them have been classified as  A-, B-, or C-type clouds.  The remaining 21 clouds could not be assigned unambiguous classes. 
More than half of the catalogued GMCs have exposed star formation (C-type) with emission peaks at several wavelengths 
within the cloud contours and these are the most massive ones. About 32$\%$ are inactive  
(A-type)  with no sources at any wavelengths and only 16$\%$ have embedded or low mass star formation (B-type) with emission peaks at MIR wavelengths only.
}

\item{The peak of the distribution of A-type clouds is near  4~kpc. Beyond 2~kpc their number density 
is comparable to that of  C-type clouds, which is the most numerous group.
Both the A-  and C-type clouds are found along HI filaments or spiral arms.  On the southern arm (but not the northern), the A-type clouds are found 
entering the arm while C-type clouds are found on the arm, suggesting the arm environment may play a role in triggering star formation. 
The average CO luminosity increases going from A- to B- to C-type clouds and this suggests that GMC mass may  continue to grow as they
evolve from the inactive phase to the formation of massive stars.
}
 
\item{We classified 611 of the 630 YSCCs into 5 categories,  b-c1-c2-c3-e-type according to the presence and location of optical,
H$\alpha$ or UV counterpart within a GMC boundary. The majority of these sources lie (in projection) within a GMC boundary, 
especially in zone~1 and zone~2. The largest class of YSCC has coincident peaks at in the UV and H$\alpha$ bands which made possible to estimate the age
and mass of the associated YSC. There is an extraordinary spatial correspondence between the GMCs and the distribution of atomic hydrogen overdensities in zones~1 and 2.
In zone~3 there are fewer GMCs, possibly because of a steepening of the molecular cloud mass spectrum with a larger fraction of clouds
being below the survey completeness limit.
}

\item{We find that GMCs classified as B- or C-type are associated with catalogued MIR-sources classified as YSCCs of b- or c-type with only a few exceptions. 
The physical association 
between GMCs and YSCCs is established in the three zones considering the whole sample of GMCs and YSCCs, independently of their prior classification. 
We analyse visually and statistically the association by generating the positional correlation function of the two distributions,  and indeed the correlation is
remarkable and stronger than with other populations in the M33 disk. If $\bar d$ is the typical separation length between YSCCs, we expect to find only 20$\%$ of the 
GMCs with a YSCC at distance less than 0.5$\bar d$ if they are randomly distributed. Instead we find fractions of order 60-70~$\%$. The correlation length is of order 
    17~pc, but there is a highly statistically significant clustering out to larger distances.   There is little or no correlation between the mass of the GMCs and that of the YSCCs.
}

\item{The extinction estimates are higher for b-type sources with weak 
or no UV/optical counterpart, which likely represent the early phases of SF. The c1-type YSCCs have visible H$\alpha$ but not FUV emission and show on 
average higher extinction than the c2-type
YSCCs, where  FUV emission is also detected. The c1-type YSCC may represent YSC at an earlier stage than c2-type  YSCC, even though the YSC age determination is not
precise enough to separate  these two classes. The most luminous YSCCs are of c2-type and are clusters which likely have completed the formation process.
}

\item{Estimated ages of most YSCCs are between 3.5 and 8~Myrs, with a marked peak around 5~Myrs, and are associated with C-type GMCs. Using the cluster ages and the
fractions of GMCs in each class i.e. in each evolutionary phase, we have estimated the GMC lifetime in M33 as being 14.2~Myrs from when they are assembled to the time when  
the YSC  breaks through the cloud, prior to gas dispersal. Even though the lifetime may be slightly longer if the cloud dispersal time is included, our estimate for the GMC lifetime 
in M33 seems  comparable to GMC lifetimes in the  Milky Way and somewhat  shorter than the  estimated GMC lifetime in the LMC.
The embedded phase, where MIR emission is visible but no H$\alpha$ or FUV emission is detected, is the shortest phase from when the cloud is 
assembled and inactive to the  switch-off of the stellar cluster formation process. 
}

\end{itemize}
 
The analysis of the largest available sample of GMCs and YSCCs and their association across the whole star forming disk of M33
provides reliable estimates of GMC lifetimes and evolutionary timescales  necessary for
for understanding the gas-star formation cycle across spiral galaxy disks.

\begin{acknowledgements}
We would like to thank the referee, Christine Wilson, for her useful comments to improve the original version of the manuscript.
\end{acknowledgements}

\vfill
\eject
 
\begin{table}
\caption{Classification and properties of the M33 molecular cloud population}
\begin{minipage}{\textwidth}
\begin{center}
\begin{tabular}{lccccccccccccc }
\hline \hline
          N$_{CO}$ &Type &RA &DEC &R$$ &r$_e$ &$\delta_{re}$ &$\sigma_v$ &$\delta_{\sigma v}$ &$\sigma_v^{gau}$ &M$_{H_2}$ &$\delta_{MH2}$ & M$_{H_2}^{vir}$ & N$_{YSCC}$ \\
            &  & &  & [kpc]& [pc]& [pc] &[km~s$^{-1}$] &[km~s$^{-1}$] & [km~s$^{-1}$] & [M$_\odot$] &[M$_\odot$] &[M$_\odot$] &\\
 \hline \hline

  1&  C&   23.36171&  30.29141&  5.7&   37&  20&   4.1&  2.5&   3.3&   0.84E+05&   0.48E+05&   0.42E+06&   2 \\            
  2&  C&   23.37007&  30.28943&  5.8&   42&  18&   4.1&  1.4&   3.7&   0.71E+05&   0.29E+05&   0.60E+06&  .. \\            
  3&  A&   23.30009&  30.38398&  4.6&   36&  14&   6.2&  2.4&   2.6&   0.19E+06&   0.30E+05&   0.26E+06&  .. \\            
  4&  A&   23.32977&  30.48902&  3.1&   56&  35&   4.7&  3.7&   2.8&   0.10E+06&   0.11E+06&   0.44E+06&  .. \\            
  5&  B&   23.32793&  30.49364&  3.1&   43&  35&   1.8&  2.0&   2.3&   0.11E+06&   0.14E+06&   0.24E+06&  87  \\           
  6&  C&   23.38505&  30.33558&  5.1&   56&  16&   8.2&  2.2&   6.1&   0.24E+06&   0.41E+05&   0.22E+07&  13,10 \\         
  7&  A&   23.43679&  30.49438&  2.7&   33&  22&   3.4&  1.6&   3.1&   0.86E+05&   0.38E+05&   0.32E+06&  .. \\            
  8&  C&   23.31071&  30.49623&  3.3&   55&  14&   4.9&  1.3&   7.1&   0.26E+06&   0.61E+05&   0.29E+07& 644 \\            
  9&  C&   23.17682&  30.37319&  6.1&  119&  13&   4.5&  0.6&   4.1&   0.90E+06&   0.40E+05&   0.21E+07& 603,601,602,600 \\

  ..& ..& ......& ......&        ...&  ...&    ...&   ..&    ..&    ..&   ........&    ........&  ........& ....\\     
\hline \hline
\end{tabular}
\end{center}
\end{minipage}
\label{tabco}
\end{table}

\begin{table}
\caption{Classification and properties of the M33 young stellar cluster candidate population}
\begin{minipage}{\textwidth}
\begin{center}
\begin{tabular}{lccccccccccccccc }
\hline \hline
          N$_{YSCC}$ &Type &N$_{CO}$ &Type &RA &DEC &lg(L$_{bol})$  &lg(L$_{TIR}$) &lg(L$_{FUV})$ &lg(L$_{H\alpha})$ &lg(M$_*$) &lg(Age) &A$_V$ & R$$ 
          &size & F$_{24}$ \\
           &  &  &  &  &  & [erg~s$^{-1}$]& [erg~s$^{-1}$] & [erg~s$^{-1}$] &[erg~s$^{-1}$] & [M$_\odot$] &[yrs] &  &[kpc]& [arcsec] & [mJy] \\
 \hline \hline

  1 & c3 &  .. & ..&   23.374977 &  30.309910 &  39.34 &  38.11 &  39.00 &  37.15 &    3.0 &    6.6 &    0.1 &    5.4 &   15.3 &   23.54  \\
  2 & c2 &   1 & C &   23.361427 &  30.290756 &  38.51 &  37.25 &  38.27 &  35.65 &    2.7 &    6.9 &    0.0 &    5.7 &    3.8 &    0.75  \\
  3 & c3 &  .. & ..&   23.372110 &  30.301689 &  39.31 &  38.78 &  38.87 &  36.84 &    3.0 &    6.7 &    0.3 &    5.5 &    7.1 &    4.57  \\
  4 & c3 &  .. & ..&   23.332920 &  30.305302 &  38.65 &  38.35 &  38.02 &  36.12 &    2.7 &    6.7 &    0.6 &    5.5 &    5.1 &    1.75  \\
  5 & e  &  .. & ..&   23.335817 &  30.304033 &  38.28 &  37.93 &  37.76 &  35.18 &    0.0 &    0.0 &    0.5 &    5.5 &    3.6 &    0.68  \\
  6 & e  &  .. & ..&   23.443571 &  30.303825 &  37.75 &  36.62 &  37.27 &  35.57 &    0.0 &    0.0 &    0.1 &    5.8 &    2.4 &    0.29  \\
  7 & e  &  .. & ..&   23.448744 &  30.304874 &  37.54 &  36.98 &  36.98 &  35.43 &    0.0 &    0.0 &    0.3 &    5.8 &    2.5 &    0.30  \\
  8 & c2 &  15 & C &   23.402447 &  30.336786 &  40.38 &  40.04 &  39.74 &  37.94 &    4.2 &    6.7 &    0.6 &    5.0 &   18.2 &   79.54  \\
  9 & c1 &  14 & D &   23.376879 &  30.319651 &  37.95 &  37.18 &  37.61 &  35.88 &    2.2 &    6.6 &    0.2 &    5.2 &    4.2 &    0.79  \\
  ..& ..& ......& ......&        ...&  ...&    ...&   ..&    ..&    ..&   ........&      ........&     .........&    ........&    ..........& ....\\     
       
\hline \hline
\end{tabular}
\end{center}
\end{minipage}
\label{tabmir}
\end{table}

\end{document}